\documentclass[times,twocolumn,final]{article}

\usepackage[margin=1in]{geometry}
\usepackage{framed,multirow}

\usepackage{amssymb}
\usepackage{latexsym}
\usepackage{gensymb}	
\usepackage{amsmath}	
\usepackage{enumitem}	
\usepackage{graphicx}	
\usepackage[square,numbers]{natbib}
\usepackage{titling}

\usepackage{url}
\usepackage{xcolor}
\definecolor{newcolor}{rgb}{.8,.349,.1}

\usepackage[draft]{hyperref} 

\usepackage[switch,pagewise]{lineno} 

\title{Body coherence in curved-space virtual reality games}

\author{Jeff Weeks}

\date{\today}

\begin{document}
\begin{titlingpage}
\maketitle
\begin{abstract}
Virtual-reality simulations of curved space are most effective and most fun when presented as a game (for example, curved-space billiards), so the user not only has something to \textit{see} in the curved space, but also has something fun to \textit{do} there.  However, such simulations encounter a geometrical problem:  they must track the player's hands as well as her head, and in curved space the effects of holonomy would quickly lead to violations of \textit{body coherence}.  That is, what the player sees with her eyes would disagree with what she feels with her hands.  This article presents a solution to the body coherence problem, as well as several other questions that arise in interactive VR simulations in curved space (radians vs. meters, visualization of the projection transformation, native-inhabitant view vs. tourist view, and mental models of curved space).
\end{abstract}
\end{titlingpage}





\onecolumn

\section{Introduction}
\label{Sec:Intro}

\begin{figure}[h!]
\centering

\begin{minipage}[c]{0.48\linewidth}
	\centering
	\includegraphics[width=0.85\linewidth]{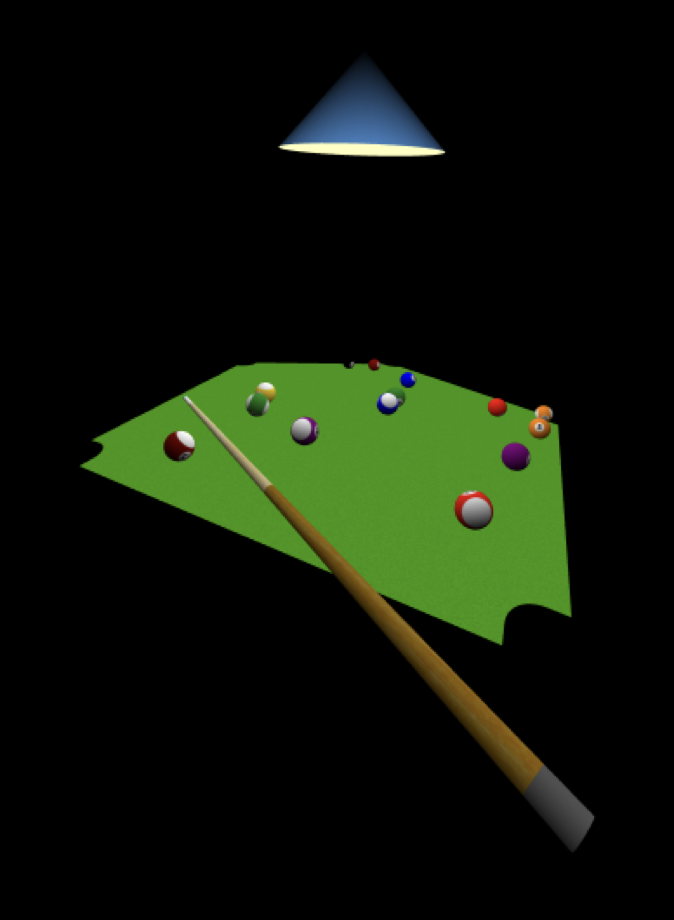}
	\caption{In hyperbolic 3-space, the billiards table is a regular pentagon with all 90\degree{} angles.}
	\label{Fig:HyperbolicBilliardsScreenshot}
\end{minipage}
\hfill	
\begin{minipage}[c]{0.48\linewidth}
	\centering
	\includegraphics[width=0.85\linewidth]{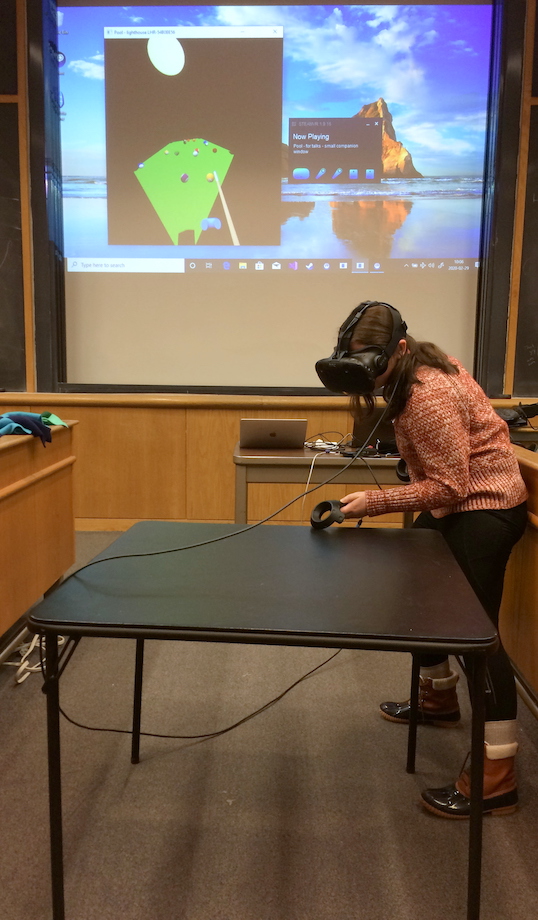}
	\caption{The hyperbolic billiards table of Figure~\ref{Fig:HyperbolicBilliardsScreenshot} agrees locally with a square Euclidean table in the lab, which adds a tactile component to the game for greater realism.}
	\label{Fig:HyperbolicBilliardsPhoto}
\end{minipage}

\end{figure}

Seeing curved spaces on a computer monitor is informative and fun, but \textit{being} in a curved space is a far richer experience and far more informative.  Indeed, for me personally, even having studied curved spaces for 45 years, when I first put on the virtual reality (VR) headset and visited a hypersphere and hyperbolic 3\nobreakdash-space, I found surprises.  And after playing a few games of billiards in those spaces (Figures \ref{Fig:HyperbolicBilliardsScreenshot} \& \ref{Fig:HyperbolicBilliardsPhoto}), I got an intuitive feel for them unlike any I had ever had before.   The reason for that deeper gut-level understanding is that VR connects not only with our conscious minds, but it also completely hijacks our subconscious understanding  of our environment \citep{Toast2017}:  it feels real!  Moreover, a billiards game is especially effective, because the need to line up a good shot forces the player to continually walk around the table and look at the ball positions and the table itself from many different viewpoints, thus building curved-space intuition more quickly.

\clearpage
\twocolumn

\subsection{Overview and related work}
\label{Sec:Overview}

Curved-space VR simulations \citep{HartEtAl2017a, Weeks2020b, Velho2020} present the player with some surprising optical effects \citep{Weeks2020b}, going beyond the surprises already found in non\nobreakdash-VR animations of the same spaces \citep{PhillipsGunn1992, Weeks2002, NovelloEtAl2020c}.  The present article explains the challenges of curved-space VR and presents some solutions for the benefit of other developers who may wish to write games or other simulations in curved space.  The most substantive---and most surprising---discovery is the need for a visitor to curved space to use her muscles to provide internal resistance to the effects of \textit{holonomy} (Section~\ref{Sec:Holonomy}), in order to maintain \textit{body coherence} (Section~\ref{Sec:BodyCoherenceProblem}).

The remainder of Section~\ref{Sec:Intro} presents some basic concepts of geometry in curved space.  Sections~\ref{Sec:BodyCoherenceProblem} and \ref{Sec:BodyCoherenceSolution} explain body coherence in detail, and propose a strategy for dealing with it in a VR simulation.  Section~\ref{Sec:CurvedSpaceGraphics} adapts the standard graphics pipeline to curved-space graphics, with new insights into the role of radians and meters, and a fresh view of the projection transformation.  Section~\ref{Sec:NativeVsTouristViews} explains how stereoscopic vision would lead a Euclidean-born tourist to grossly misjudge distances in curved space.  That same section then goes on to recommend that curved-space VR games be written to simulate distances as each space's native-born inhabitants perceive them, and shows how to modify the graphics pipeline to achieve that goal.  Section~\ref{Sec:MentalModels} notes that how a player perceives a curved space depends not only on what the player's eyes see, but also on how the player's brain integrates that visual data into a mental model of the space.  Finally, Section~\ref{Sec:ConclusionsAndFutureWork} lays out some directions for future work.

This project as a whole takes its inspiration from, and builds upon, the pioneering work of \citep{HartEtAl2017a}.  We restrict our attention to the three isotropic spaces: spherical, Euclidean, and hyperbolic.  Other authors have produced beautiful animations of non\nobreakdash-isotropic spaces, in both VR \citep{HartEtAl2017b, CoulonEtAl2020a, CoulonEtAl2020b, CoulonEtAl2020c, NovelloEtAl2020b} and traditional graphics \citep{Weeks2006, Berger2015, KopczynskiCelinska2019, KopczynskiCelinska2020, NovelloEtAl2020a}.  For an excellent overview of the existing literature, see Section~1.3 of \citep{CoulonEtAl2020c}.

Note that some well-known games start with several Euclidean spaces and then use portals to make ad~hoc connections between them.  By contrast, our billiards game is played in a single self-consistent non-Euclidean space (the player may choose spherical, Euclidean or hyperbolic), with no need for portals.

The author thanks the anonymous referees for their excellent suggestions that greatly improved the quality of this article.

\subsection{Definitions and notation}
\label{Sec:Terminology}

Just as a circle (or 1\nobreakdash-sphere) is the boundary of a disk, and an ordinary sphere (or 2\nobreakdash-sphere) is the surface of a solid ball, a \textit{hypersphere}, or \textit{3\nobreakdash-sphere}, is the 3\nobreakdash-dimensional hypersurface of a 4\nobreakdash-dimensional hyperball.  Because the 3\nobreakdash-sphere is intrinsically 3- (not 4-) dimensional, it represents a possible universe for creatures like us---and also a fun place to play billiards and other games!

\begin{figure}[ht!]
\centering
\centerline{\includegraphics[width=0.9\linewidth]{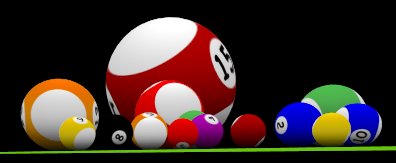}}
\caption{In a 3\nobreakdash-sphere, if the player takes a step back from the table, the nearest balls look smaller but more distant balls look larger.}
\label{Fig:SphericalOptics}
\end{figure}

\begin{figure}[b!]
\centering
\centerline{\includegraphics[width=0.85\linewidth]{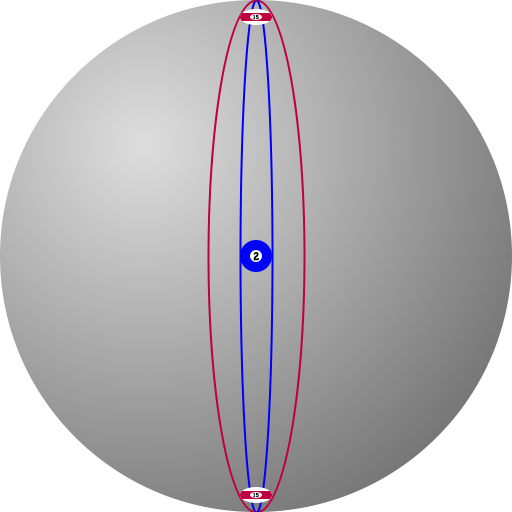}}
\caption{A ball near the antipodal point subtends the same angle as a nearby ball.}
\label{Fig:GeodesicConvergence}
\end{figure}

For example, a surprising optical effect in the 3\nobreakdash-sphere is that a billiard ball looks smallest when it's 90\degree{} away from the observer.  If it gets any further away than 90\degree{} it starts to look larger again (Figure~\ref{Fig:SphericalOptics}), because a distant ball sitting $(180 - \theta)\degree{}$ away from the observer subtends the same angle in the observer's field of view that a nearby ball sitting $\theta\degree{}$ away would subtend (Figure~\ref{Fig:GeodesicConvergence}).  A ball sitting at the observer's antipodal point (180\degree{} away) would fill the whole sky!

\newpage

\begin{figure}[t!]
\centering
\includegraphics[width=0.75\linewidth]{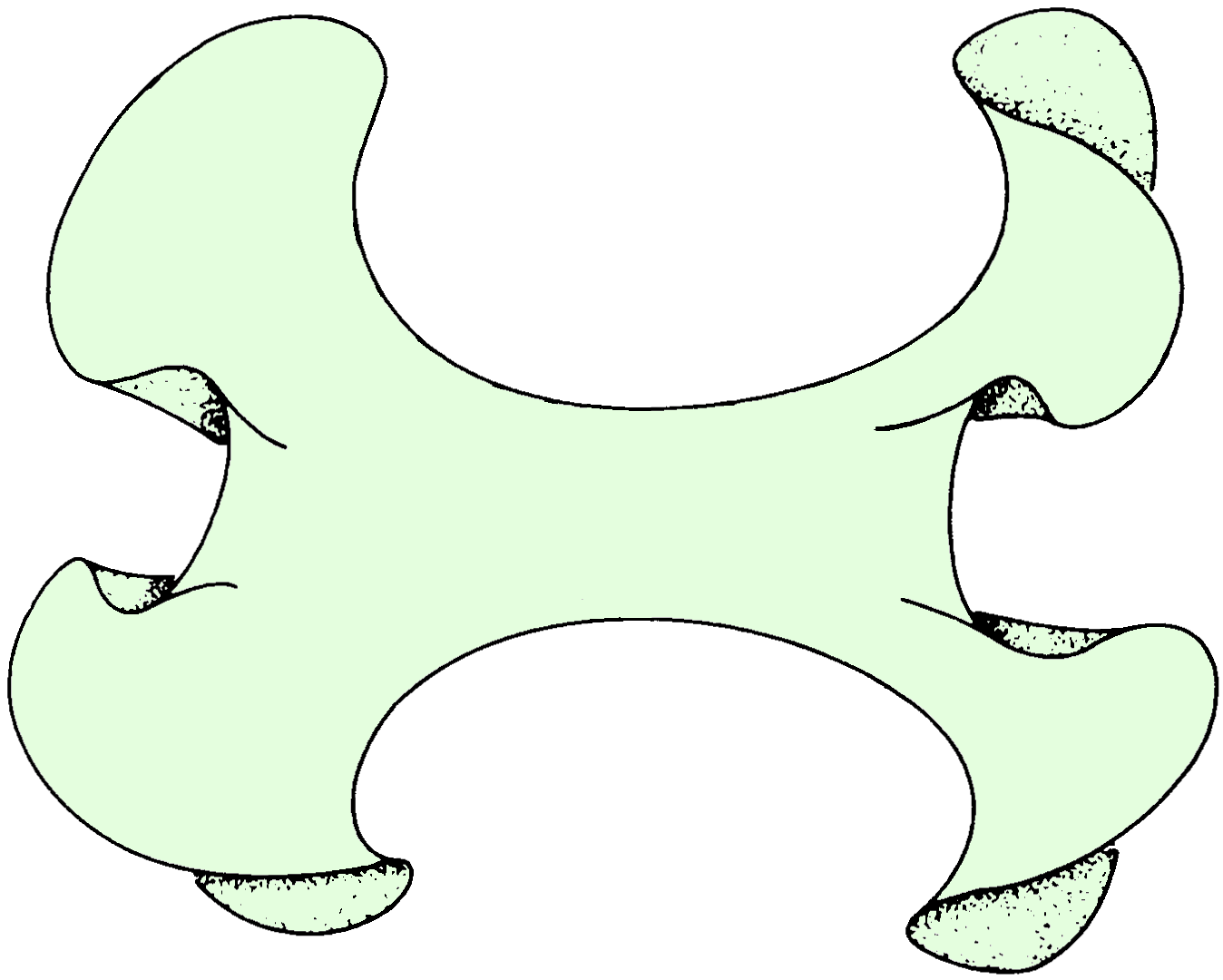}
\caption{When sitting in Euclidean space, a piece of a hyperbolic plane is very curly.}
\label{Fig:HyperbolicPlane}
\end{figure}

\begin{figure}[b!]
\centering
\includegraphics[width=0.85\linewidth]{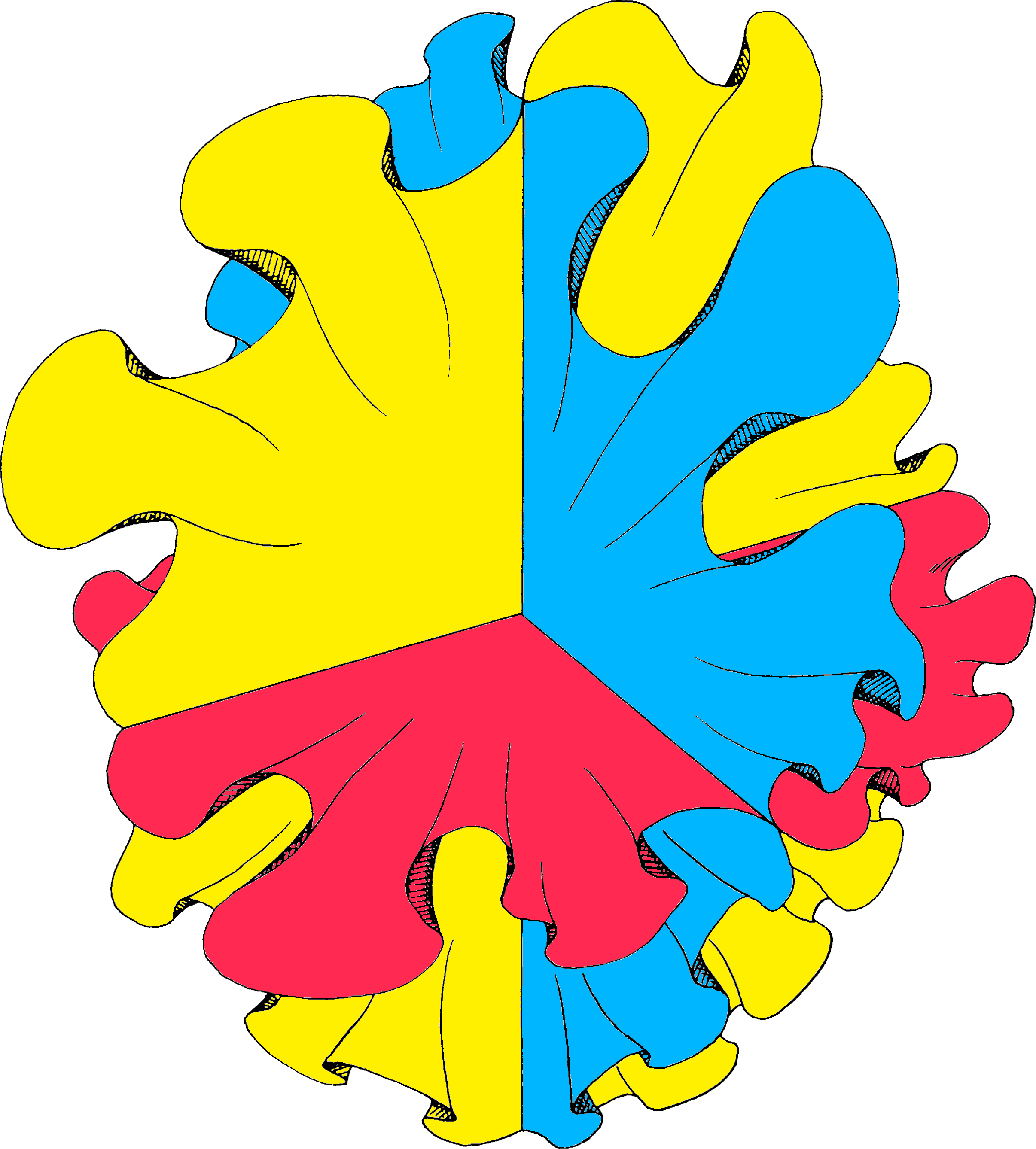}
\caption{An ``artist's conception'' of the fact that every slice of hyperbolic 3-space is a hyperbolic plane.}
\label{Fig:HyperbolicSpace}
\end{figure}

The \textit{hyperbolic plane} is intrinsically like a curly piece of leaf lettuce (Figure~\ref{Fig:HyperbolicPlane}).  \textit{Hyperbolic 3\nobreakdash-space} is similar, but one dimension higher (Figure~\ref{Fig:HyperbolicSpace}). For brevity, one often lets
\[
\begin{array}{ll}
	\mathrm{S}^3 & \text{denote the 3\nobreakdash-sphere,}			\\
	\mathrm{E}^3 & \text{denote Euclidean 3\nobreakdash-space, and}	\\
	\mathrm{H}^3 & \text{denote hyperbolic 3\nobreakdash-space.}
\end{array}
\]

\begin{figure}[t!]
	\centering
	\includegraphics[width=\linewidth]{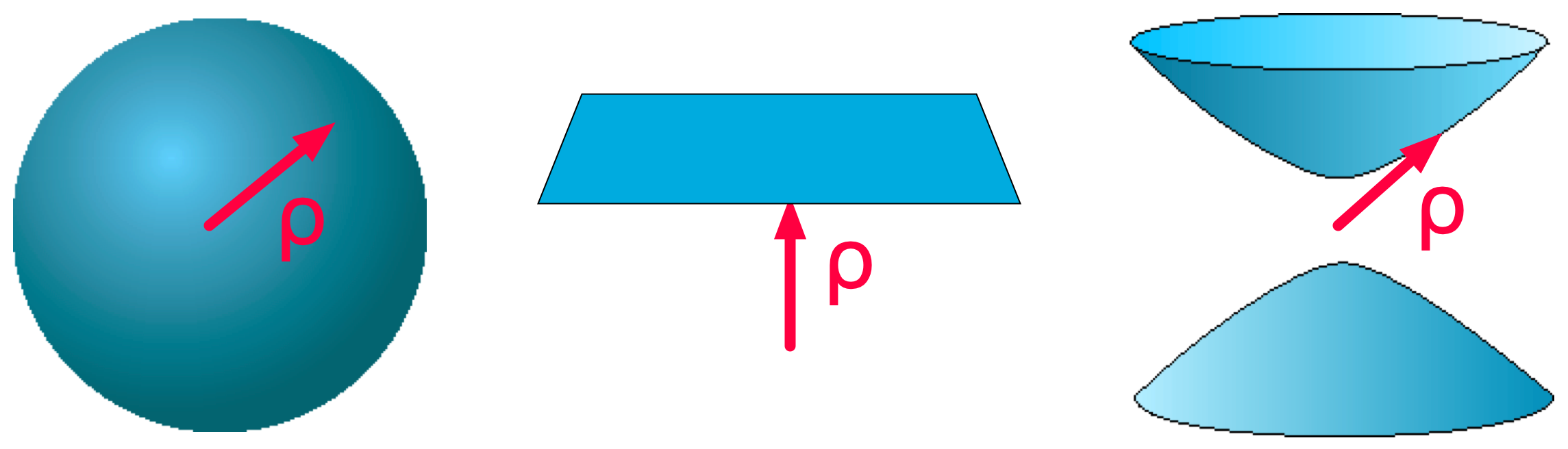}
	\caption{The concept of \textit{radius} applies not only to spheres \textit{(left)}, but also to hyperbolic spaces \textit{(right)} and even Euclidean spaces \textit{(center)}.}
	\label{Fig:Radii}
\end{figure}

A 3\nobreakdash-sphere $\mathrm{S}^3$ of radius $\rho$ may be formally defined as the set of points $\rho$ units from the origin in Euclidean 4\nobreakdash-space, where the radius $\rho$ is measured using the \textit{Euclidean metric}
\[
	\lvert(x,y,z,w)\rvert^2 = x^2 + y^2 + z^2 + w^2 = \rho^2.
\]

\noindent By contrast, a hyperbolic 3\nobreakdash-space of radius $\rho$ is most naturally defined not in Euclidean 4\nobreakdash-space, but in Minkowski space, as the set of points $\rho$ units from the origin, where the radius $\rho$ is measured using the \textit{Lorentz metric}
\[
	\lvert(x,y,z,w)\rvert^2 = x^2 + y^2 + z^2 - w^2 = -\rho^2.
\]
For an elementary yet thorough introduction to this definition of hyperbolic 3\nobreakdash-space, please see the section ``What's hyperbolic space?'' in \citep{Weeks2002}.  Typically one works with the upper sheet of the hyperboloid ($w > 0$) and ignores the lower half ($w < 0$).  Figure~\ref{Fig:Radii} of the present article illustrates these definitions, but one dimension lower, showing $\mathrm{S}^2$, $\mathrm{E}^2$ and $\mathrm{H}^2$ instead of $\mathrm{S}^3$, $\mathrm{E}^3$ and $\mathrm{H}^3$.  In Section~\ref{Subsec:Radians} it will be convenient to define the Euclidean space of radius $\rho$ to be the hyperplane $w = \rho$.

Rigid motions of $\mathrm{S}^3$ are represented by \textit{rotation matrices} from the \textit{rotation group} $\mathrm{O}(4)$.  Analogously, rigid motions of $\mathrm{H}^3$ are represented by \textit{Lorentz matrices} from the \textit{Lorentz group} $\mathrm{O}(3,1)$.  Readers who wish to learn more about these matrices may find an elementary explanation in \citep{Weeks2002}, but that explanation isn't required to understand the present article.

\subsection{Construction of the billiards tables}
\label{Sec:BilliardsTables}

\begin{figure}[t!]
\centering

\centerline{\includegraphics[width=0.48\linewidth]{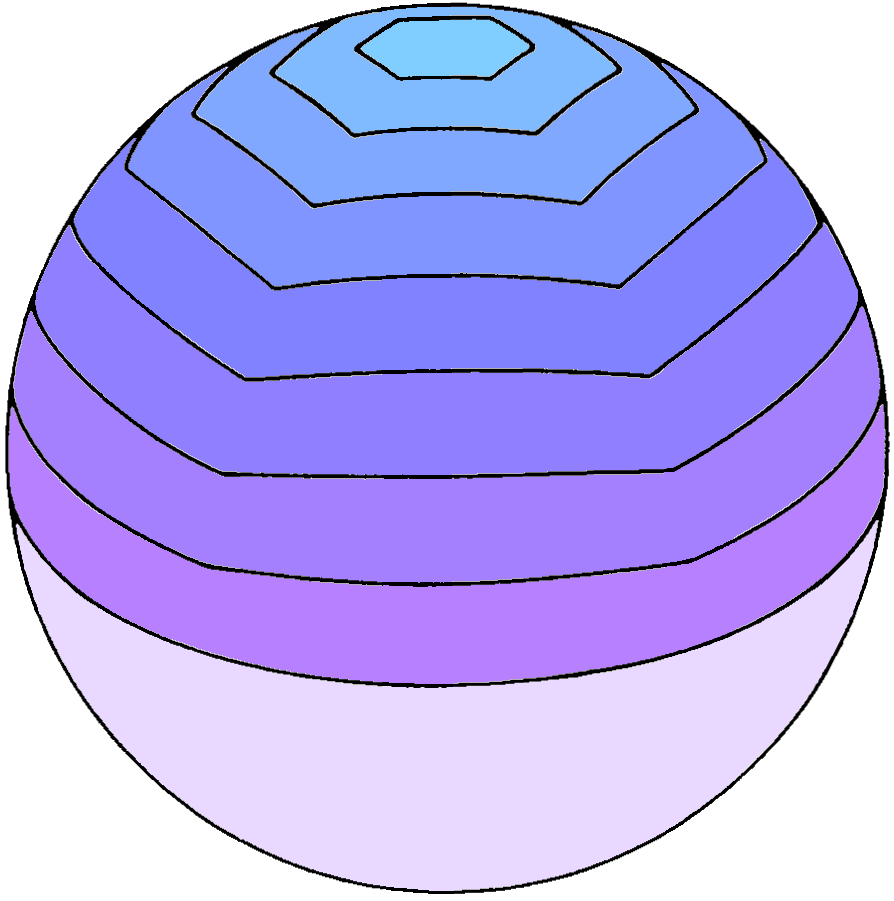}}
\centerline{\footnotesize \textit{(a) In spherical geometry, 
  larger \textit{n}-gons have broader angles.}}

\bigskip

\centerline{\includegraphics[width=0.72\linewidth]{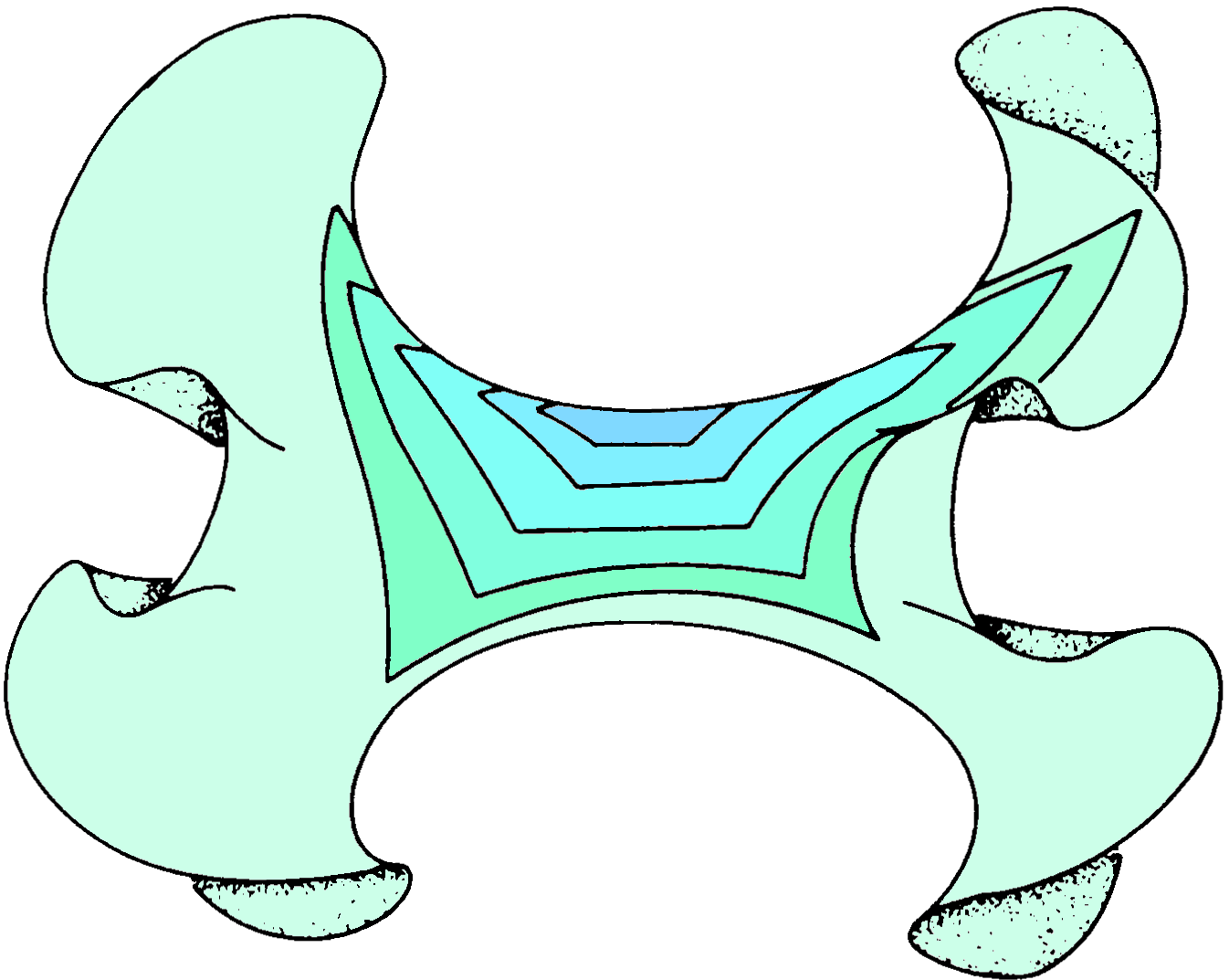}}
\centerline{\footnotesize \textit{(b) In hyperbolic geometry, 
  larger \textit{n}-gons have sharper angles.}}

\caption{A polygon's angles depend on the polygon's area.}
\label{Fig:Polygons}
\end{figure}


\begin{figure}[t!]
\centering

\includegraphics[width=\columnwidth]{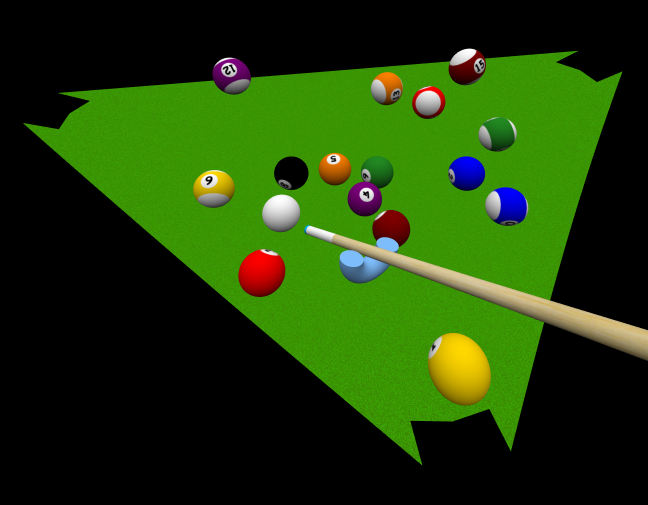}
\centerline{\footnotesize \textit{(a) Typical player's viewpoint}}

\vspace{0.05\columnwidth}

\includegraphics[width=0.779321\columnwidth]{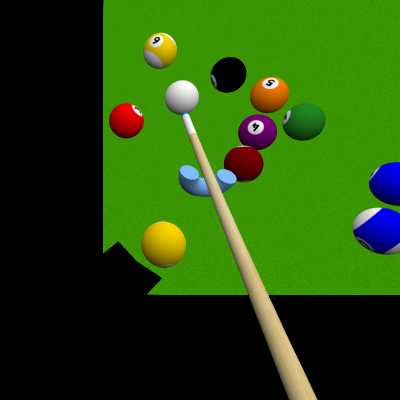}
\centerline{\footnotesize \textit{(b) Seen from above a corner}}

\caption{The billiards table in $\mathrm{S}^3$ is an equilateral right triangle.}
\label{Fig:TriangularTable}
\end{figure}


In each of the three spaces, we want to construct a billiards table
\begin{itemize}[noitemsep,topsep=0pt]
\item whose sides all have the same length, and
\item whose corners all have 90\degree{} angles,
\end{itemize}
so that the virtual table in the curved space will at least locally match the player's square physical table in the lab (Figure~\ref{Fig:HyperbolicBilliardsPhoto}).  In Euclidean 3\nobreakdash-space this is trivially easy:  we use a square virtual table whose dimensions match those of the physical table.

In the 3\nobreakdash-sphere $\mathrm{S}^3$, no right-angled square is possible, because the angles of a regular $n$\nobreakdash-gon in spherical geometry are always at least slightly greater than the angles of the corresponding Euclidean $n$\nobreakdash-gon.  More precisely, the angles of a very small $n$\nobreakdash-gon are close to the Euclidean value, but then as we let the $n$\nobreakdash-gon grow, its angles get wider and wider (Figure~\ref{Fig:Polygons}(a)).  To construct the desired billiards table, we start not with a square but with an equilateral triangle.  A small equilateral triangle on a sphere has angles close to the Euclidean value of 60\degree{}, so we let the triangle grow until its angles reach 90\degree{}, and use that 90\degree{} equilateral triangle for our billiards table in $\mathrm{S}^3$ (Figure~\ref{Fig:TriangularTable}).

In hyperbolic 3\nobreakdash-space $\mathrm{H}^3$, by contrast, the angles of a regular $n$\nobreakdash-gon are always less than the Euclidean values, with larger $n$\nobreakdash-gons having progressively smaller angles (Figure~\ref{Fig:Polygons}(b)).  So to construct our right-angled billiards table, we start with a tiny regular pentagon, whose angles are close to the Euclidean value of 108\degree{}, let it grow until its angles shrink to 90\degree{}, and use that 90\degree{} regular pentagon for our billiards table in $\mathrm{H}^3$ (Figures \ref{Fig:HyperbolicBilliardsScreenshot} and \ref{Fig:TableEdgeOn}).  (Alternatively any regular $n$\nobreakdash-gon, with $n \geq 5$, would work.)

The image of the square tabletop in Figure~\ref{Fig:HyperbolicBilliardsPhoto}, when considered as a 2\nobreakdash-dimensional photo on the page, has corner angles that are significantly different from 90\degree{}.  Yet we know from experience that those corner angles really are 90\degree{}, and would look like 90\degree{} angles when viewed from directly above.  Similarly, our triangular billiards table has corner angles that look significantly different from 90\degree{} when viewed obliquely (Figure~\ref{Fig:TriangularTable}(a)), yet they really are 90\degree{}, and look like 90\degree{} angles when viewed from directly above (Figure~\ref{Fig:TriangularTable}(b)).  And, of course, the same is true for the 90\degree{} angles of our pentagonal billiards table in $\mathrm{H}^3$.

\subsection{Straight lines look straight}
\label{Sec:StraightLines}

\begin{figure}[b!]
\centering
\includegraphics[width=\linewidth]{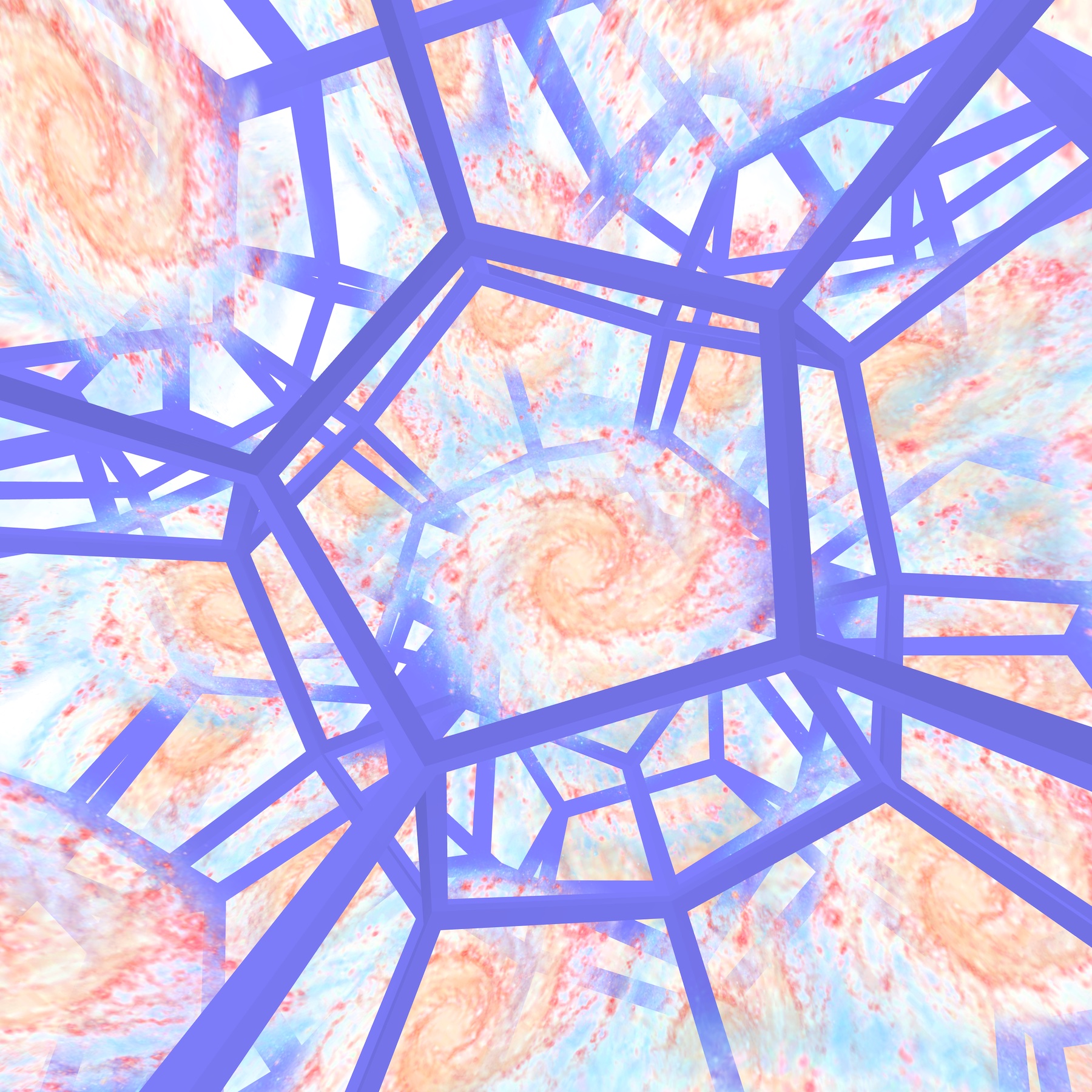}
\caption{A tiling of the 3-sphere by dodecahedra with 120\degree{} edge angles.}
\label{Fig:DodecahedralTilingSph}
\end{figure}

\begin{figure}[b!]
\centering
\includegraphics[width=\linewidth]{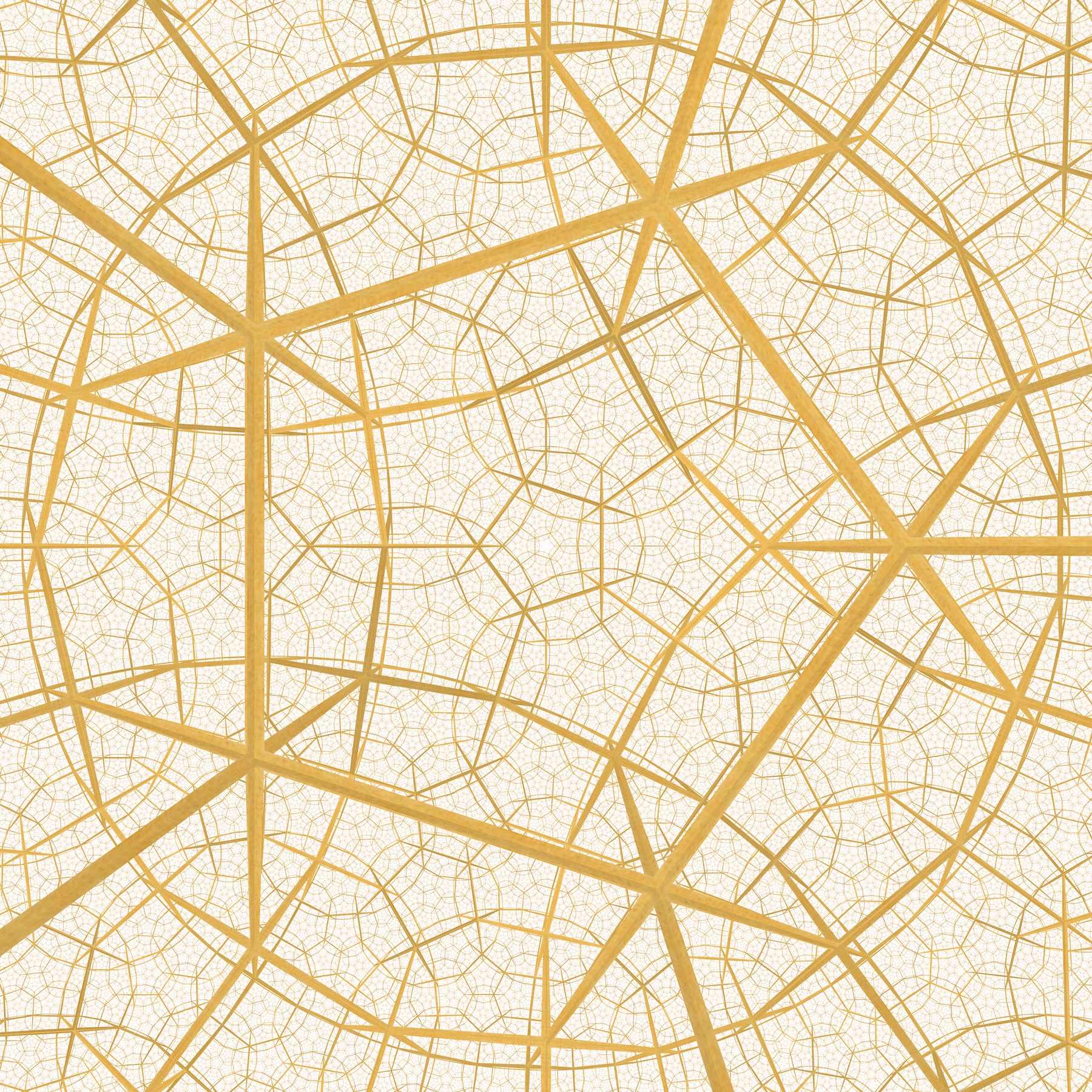}
\caption{A tiling of hyperbolic 3-space by dodecahedra with 90\degree{} edge angles.}
\label{Fig:DodecahedralTilingHyp}
\end{figure}

Figure~\ref{Fig:DodecahedralTilingSph} shows a tiling of a 3\nobreakdash-sphere by dodecahedra with edge angles of 120\degree{}, which is slightly more than the roughly 117\degree{} angles of a regular dodecahedron in flat space.
Figure~\ref{Fig:DodecahedralTilingHyp} shows a tiling of hyperbolic 3\nobreakdash-space by dodecahedra with edge angles of 90\degree{} which, by contrast, is less than the angles of a flat-space dodecahedron.  But please note that in both cases, even though the space is curved, \textit{straight lines all look straight}, in sharp contrast to the popular conception of ``curved space'' as a place where everything looks wavy.  That popular conception would be valid in a non-homogeneous or non-isotropic space \citep{Berger2015, HartEtAl2017b, CoulonEtAl2020a, CoulonEtAl2020b, CoulonEtAl2020c, KopczynskiCelinska2019, KopczynskiCelinska2020, NovelloEtAl2020a, NovelloEtAl2020b, Weeks2006}, but the 3\nobreakdash-sphere and hyperbolic 3\nobreakdash-space are both homogeneous and isotropic.  Of course, even though all straight lines look straight, they behave differently than in flat space:  only in hyperbolic 3\nobreakdash-space can you have a pentagonal billiards table with 90\degree{} corners (Figures \ref{Fig:HyperbolicBilliardsScreenshot} and \ref{Fig:TableEdgeOn}), and only in a 3\nobreakdash-sphere can you have a triangular table with 90\degree{} corners (Figure~\ref{Fig:TriangularTable}).

\subsection{Billiards table is planar}
\label{Sec:PlanarTable}

\begin{figure}[b!]
\centering

\includegraphics[width=0.55\columnwidth]{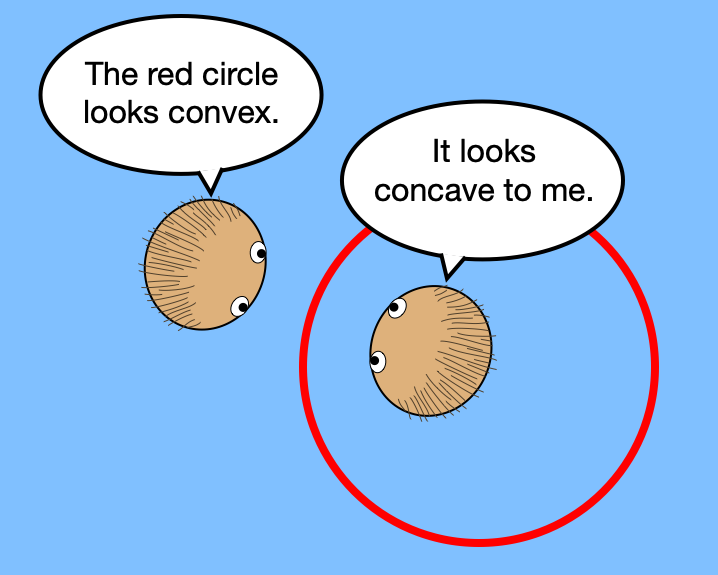}
\centerline{\footnotesize \textit{(a) A circle in a Euclidean plane}}
\smallskip

\includegraphics[width=0.5\columnwidth]{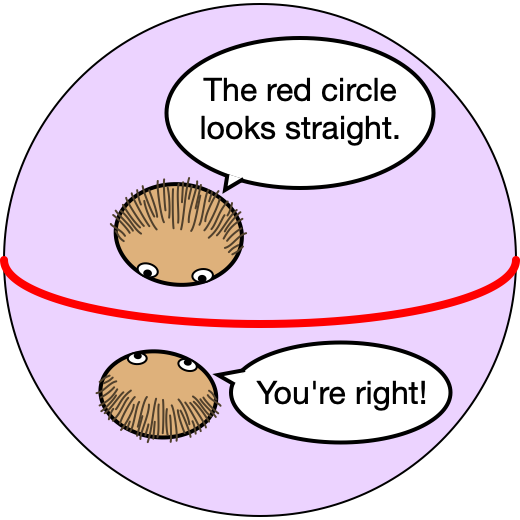}
\centerline{\footnotesize \textit{(b) A great circle on a sphere}}
\smallskip

\includegraphics[width=0.65\columnwidth]{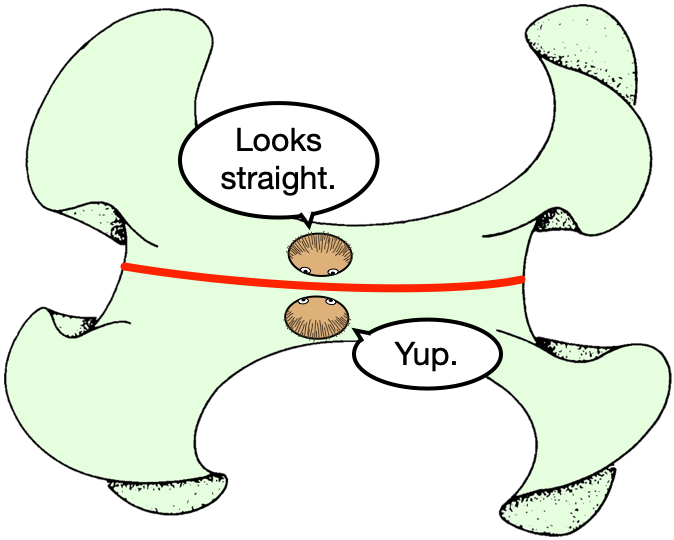}
\centerline{\footnotesize \textit{(c) A line in a hyperbolic plane}}
\smallskip

\caption{Whether a circle looks convex/concave or straight depends on the surface that it sits in.}
\label{Fig:Circles}
\end{figure}

You might be wondering why the billiards table in Figure~\ref{Fig:HyperbolicBilliardsScreenshot}, even though it's hyperbolic and has five 90\degree{} corners, nevertheless looks flat (not at all curly).  To understand why, first consider how 2\nobreakdash-dimensional beings would see a circle.  A circle drawn in the Euclidean plane (Figure~\ref{Fig:Circles}(a)) looks convex to an observer viewing it from the outside, but concave to an observer viewing it from the inside. By contrast, a sphere's equator (Figure~\ref{Fig:Circles}(b)) is a circle that looks perfectly straight---neither convex nor concave---to observers on both sides.  Moving up a dimension, an ordinary sphere drawn in Euclidean space looks convex to an observer viewing it from the outside, but concave to an observer viewing it from the inside. By contrast, the ``equator'' of a 3\nobreakdash-sphere is a 2\nobreakdash-sphere that looks perfectly planar---neither concave nor convex---to observers on both sides.  That's why the faces of the dodecahedra in Figure~\ref{Fig:DodecahedralTilingSph} all look planar---neither concave nor convex---even though they sit in a 3\nobreakdash-sphere.

\begin{figure}[b!]
	\centering
	\begin{minipage}[b]{\columnwidth} 

		\centerline{\includegraphics[width=\linewidth]{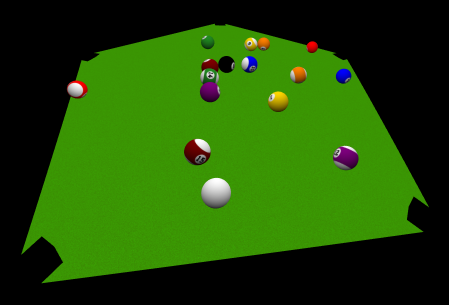}}
		{
			\footnotesize 
			\centerline{\footnotesize \textit{(a) A typical game in progress}}
			\centerline{\footnotesize \textit{on a hyperbolic billiards table.}}
		}
		\medskip
		\centerline{\includegraphics[width=\linewidth]{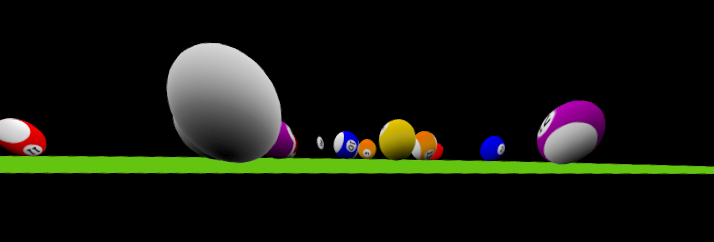}}
		{
			\footnotesize 
			\centerline{\footnotesize \textit{(b) Viewing the same game at eye level confirms}}
			\centerline{\footnotesize \textit{that the hyperbolic table is perfectly planar.}}
		}
	
		\caption{Two views of the same hyperbolic billiards table.}

	\label{Fig:TableEdgeOn}
	\end{minipage}
\end{figure}

Similarly, a slice of a hyperbolic plane is a line that looks straight to observers on both sides (Figure~\ref{Fig:Circles}(c)), and a slice of hyperbolic 3\nobreakdash-space is a hyperbolic plane that looks perfectly straight---not at all saddle-shaped---to observers on both sides.  That's why the faces of the dodecahedra in Figure~\ref{Fig:DodecahedralTilingHyp} all look planar.  Any one of those faces may be chosen to serve as the right-angled, pentagonal, hyperbolic---yet perfectly planar---billiards table shown in Figure~\ref{Fig:HyperbolicBilliardsScreenshot}.  Figure~\ref{Fig:TableEdgeOn}(a) shows a typical billiards game in progress in $\mathrm{H}^3$;  Figure~\ref{Fig:TableEdgeOn}(b) shows the same game with the table viewed from approximately eye level, to confirm that the hyperbolic table really is planar (not saddle-shaped).

\subsection{Holonomy}
\label{Sec:Holonomy}

\begin{figure}[b!]
\centering
\includegraphics[width=0.5\columnwidth]{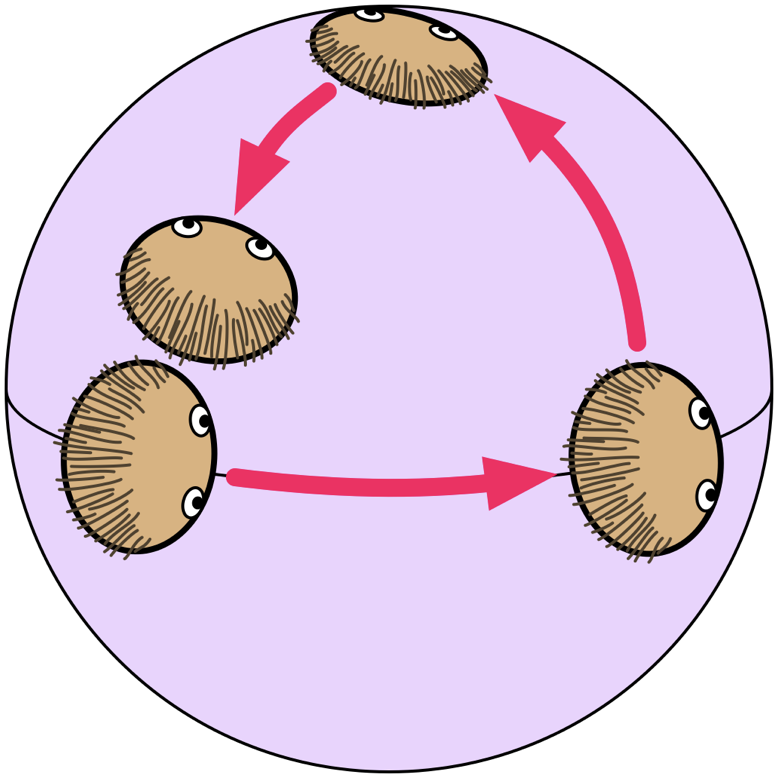}
\centerline{\footnotesize \textit{(a) Positive holonomy}}
\smallskip

\includegraphics[width=0.55\columnwidth]{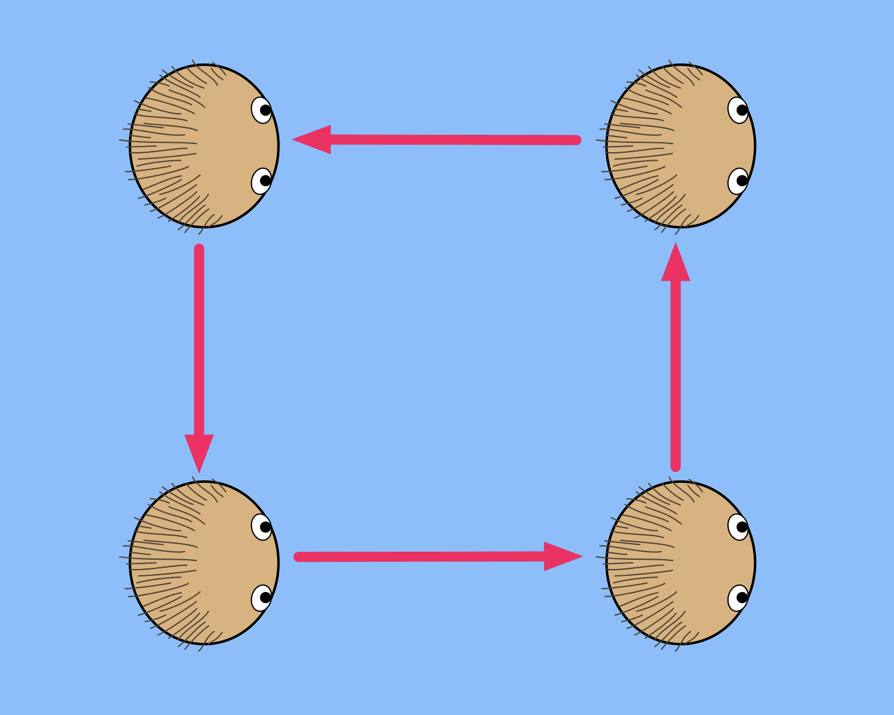}
\centerline{\footnotesize \textit{(b) Zero holonomy}}
\smallskip

\includegraphics[width=0.85\columnwidth]{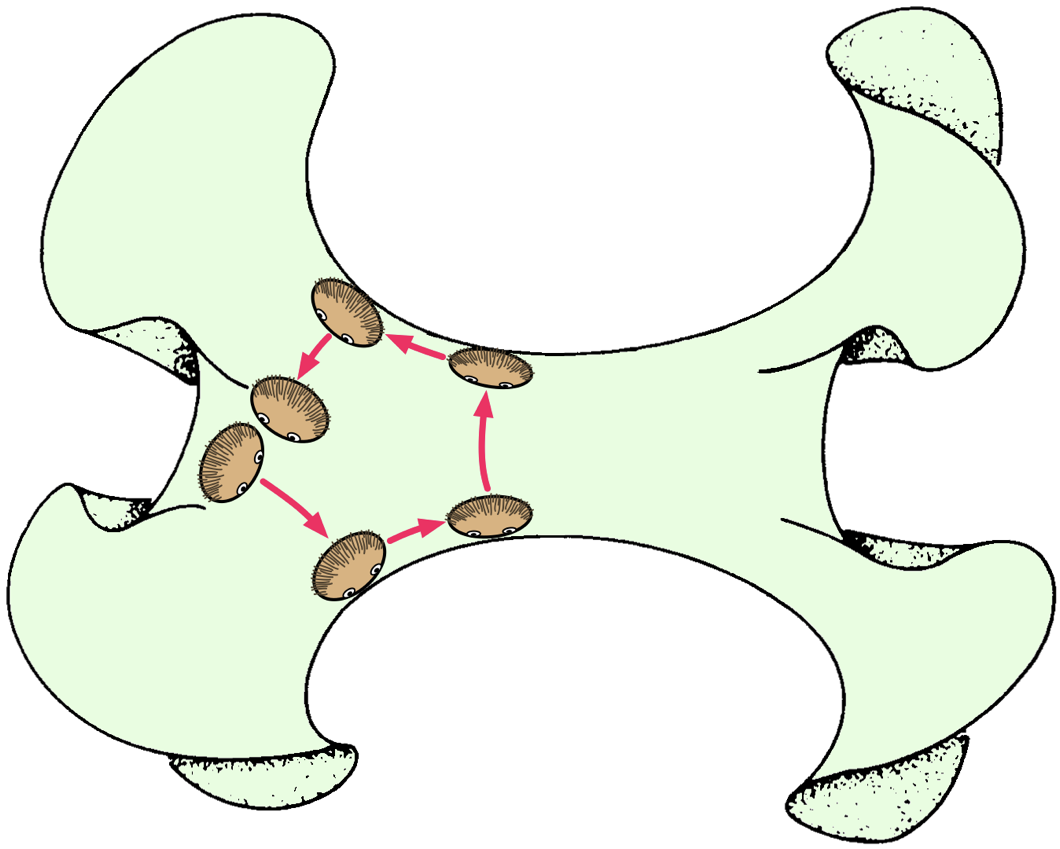}
\centerline{\footnotesize \textit{(c) Negative holonomy}}

\caption{Holonomy}
\label{Fig:HolonomyIntro}
\end{figure}

As explained in \citep{Weeks2020b}, an observer who slides around a loop in Euclidean space (being careful not to spin on her own axis as she slides around!) comes back unrotated (Figure~\ref{Fig:HolonomyIntro}(b)).  But an observer who slides around a loop on a sphere (again taking care not to spin on her own axis) comes back rotated relative to how she started (Figure~\ref{Fig:HolonomyIntro}(a)).  This effect is called \textit{positive holonomy}.  For brevity, sliding-without-spinning-on-one's-own-axis is called \textit{parallel translation}.  An observer who parallel-translates around a loop in hyperbolic space also comes back rotated (Figure~\ref{Fig:HolonomyIntro}(c)), but in the opposite sense (clockwise or counterclockwise) to the sense in which she slid around the loop.  This is called \textit{negative holonomy}.

On a homogeneous surface, the exact angle $\theta$ by which the observer gets rotated is proportional to the area $A$ that her loop encloses (for an elementary proof, one may imitate the methods in Chapter~12 of \citep{Weeks2020a}).  Because the holonomy angle $\theta$ is proportional to the enclosed area $A$, we may talk about the density $k$ of holonomy per unit area, in symbols
$$
  k \equiv \frac{\theta}{A}.
$$
While there are many possible definitions of curvature---all ultimately equivalent to each other---a particularly simple one is to define a homogeneous surface's \textit{intrinsic (or Gaussian) curvature} to be this density $k$ of holonomy per unit area, where the holonomy angle is measured in radians (not degrees).
 
A 3\nobreakdash-dimensional space is \textit{homogeneous} if its geometry is the same at all points.  A homogeneous space is, in addition, \textit{isotropic} if all its 2\nobreakdash-dimensional slices have the same intrinsic curvature $k$.  By contrast, it's \textit{anisotropic} if different 2\nobreakdash-dimensional slices sometimes have different intrinsic curvatures.  For example, in an anisotropic space, the ``vertical'' slices might have zero curvature while the ``horizontal'' slices have positive curvature.  In the present article, we consider only the three isotropic spaces $\mathrm{S}^3$, $\mathrm{E}^3$ and $\mathrm{H}^3$.

\section{Headset tracking and body coherence: the problem}
\label{Sec:BodyCoherenceProblem}

When simulating a curved space in VR, a fundamental question is how to map a pose of the user's headset in the physical lab to a pose of the user's virtual self in the curved space.  The simplest algorithm would be to map local motions of the headset (as measured in its own local coordinate system) to the same local motions of the user's virtual head (as measured in its own local coordinate system).  In other words, if the physical headset moves 1\,cm to its left in the lab, the user's virtual head moves 1\,cm to its left in the curved virtual space;  if the physical headset rotates 2\degree{}, the user's virtual head rotates 2\degree{}; and so on.

\begin{figure}[t!]
	\centering
	\includegraphics[width=0.60\linewidth]{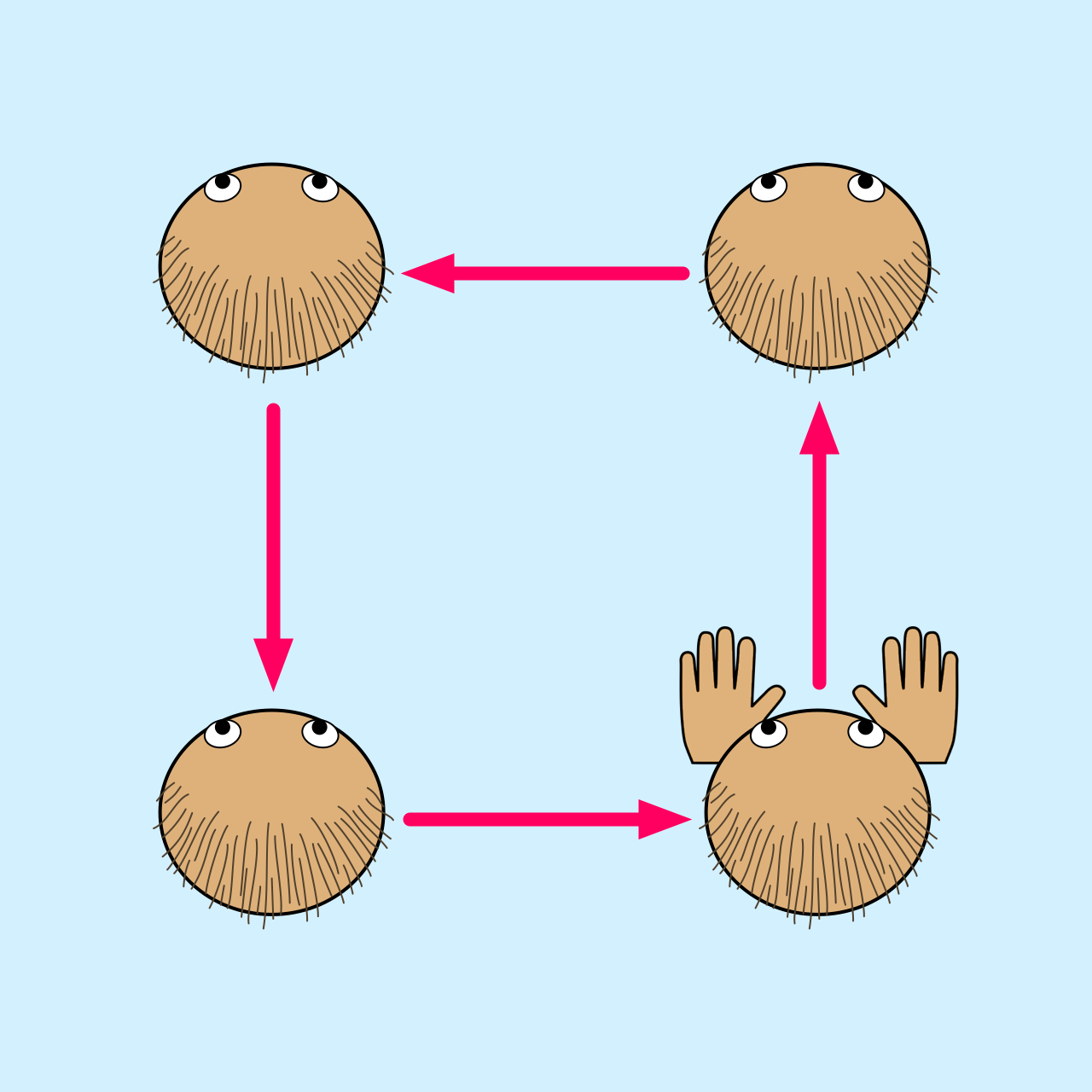}
	\caption{The player moves her head around a small square
		in the physical Euclidean lab, while leaving her body
		and hands still.}
	\label{Fig:HolonomyLab}
\end{figure}


\begin{figure*}[t!]
	\centering
	\begin{minipage}[b]{0.60\columnwidth} 
		\centerline{\includegraphics[width=\linewidth]{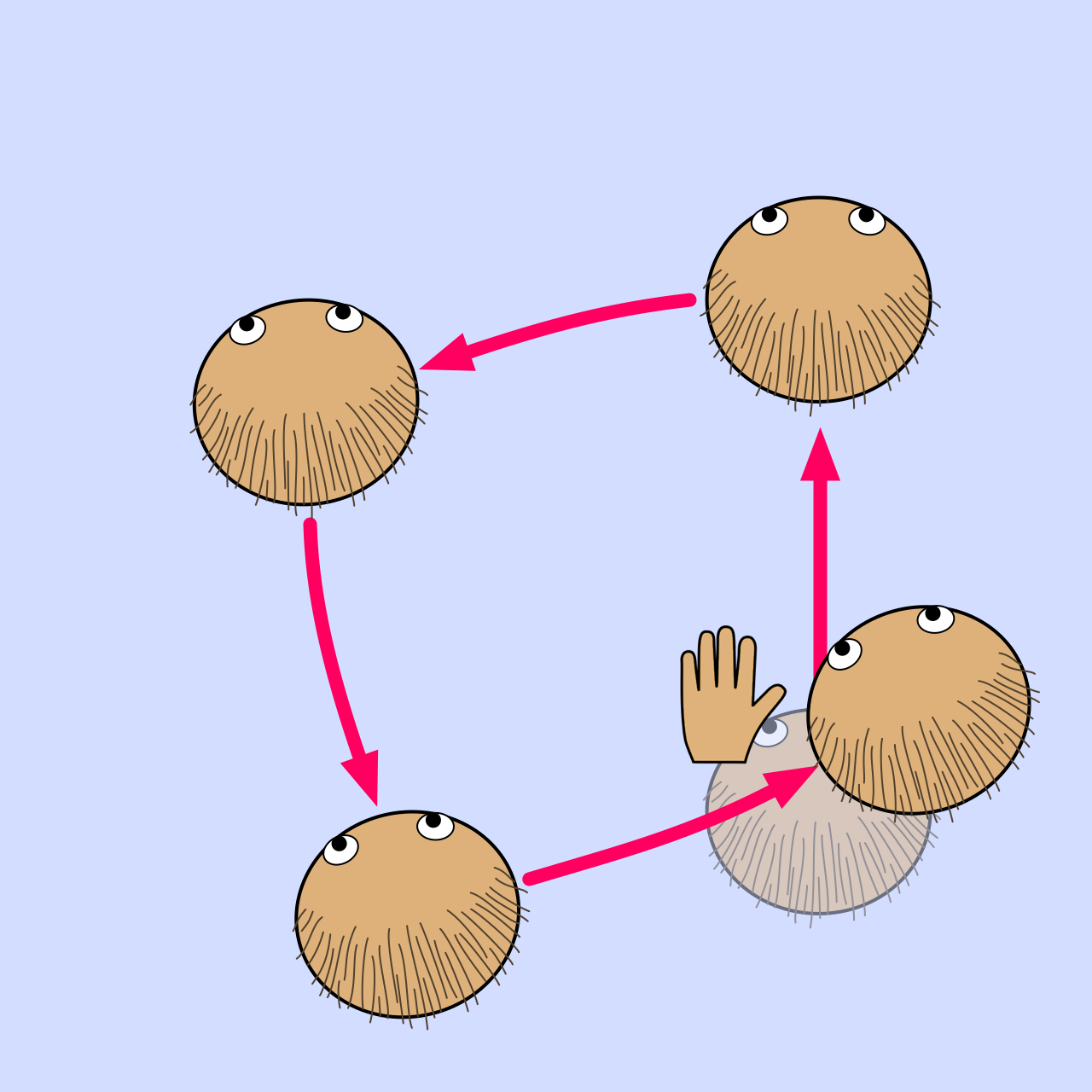}}
		\centerline{\footnotesize \textit{(a) Positive holonomy in a 3-sphere}}
	\end{minipage}
	\hspace{0.04\linewidth}
	\begin{minipage}[b]{0.60\columnwidth} 
		\centerline{\includegraphics[width=\linewidth]{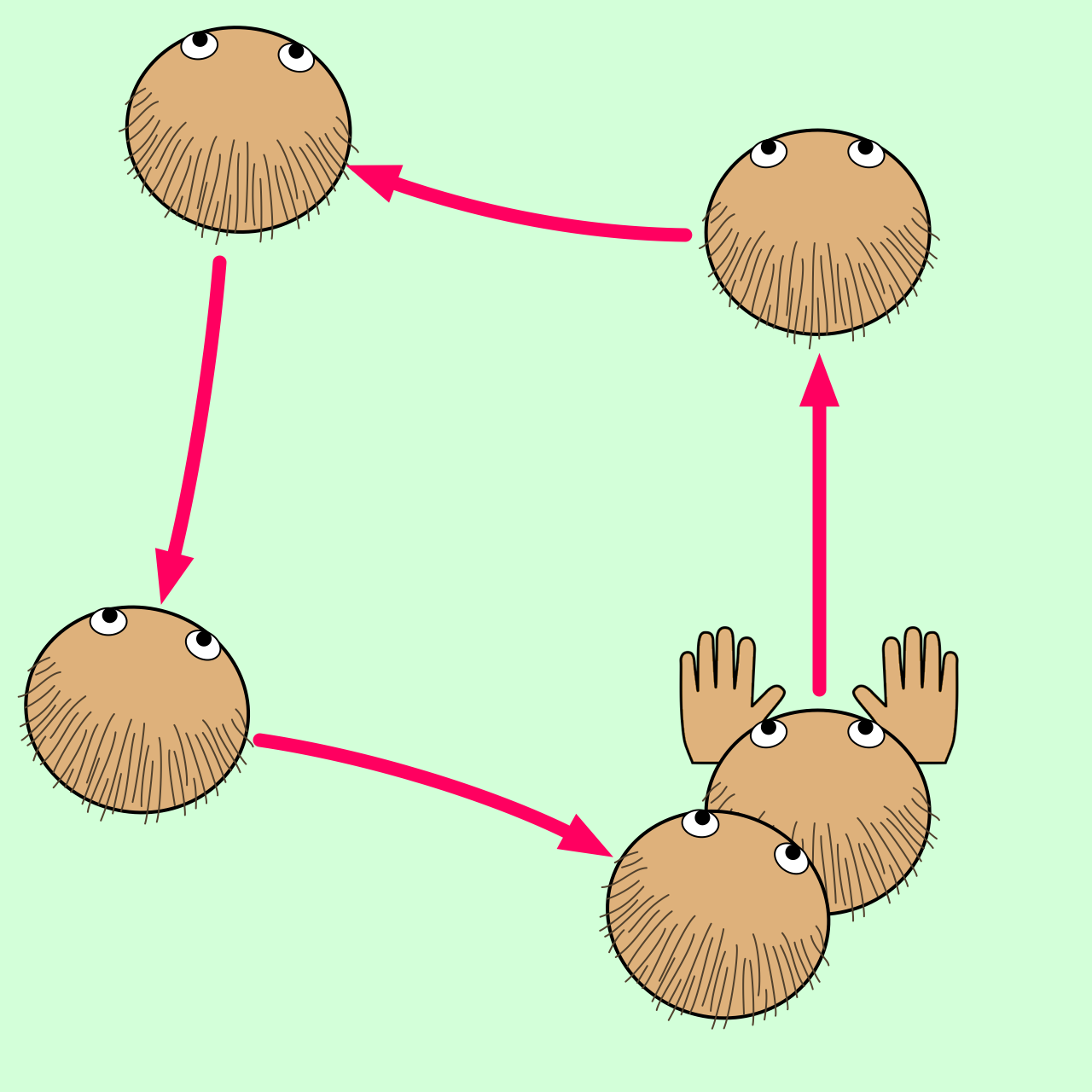}}
		\centerline{\footnotesize \textit{(b) Negative holonomy in hyperbolic 3-space}}
	\end{minipage}
	\caption{In a curved space, the player's virtual head
		ends up offset and rotated, by an amount proportional
		to the area that her path encloses.}
	\label{Fig:HolonomySim}
\end{figure*}

That naive algorithm would work fine if we were tracking only the user's head.  But in practice we must track the user's two hands as well, so that, for example, a player in the curved-space billiards game can take a shot.  Unfortunately holonomy (defined in Section~\ref{Sec:Holonomy} and illustrated in Figure~\ref{Fig:HolonomyIntro}) can force the player's head and hands out of alignment when using the naive tracking algorithm.  To see how, consider what happens if the player parallel-translates her head a small distance $d$ forward, then the same distance to her left, the same distance backwards, and finally the same distance to her right.  In the physical Euclidean lab, her head returns to its original position and orientation (Figure~\ref{Fig:HolonomyLab}).  But in the 3\nobreakdash-sphere, those forward, leftward, backward, and rightward motions are realized not by Euclidean translations, but by rotations of the 3\nobreakdash-sphere.  If we place the player at the north pole $(0,0,0,1)$ and let $\theta = d/R$, where $R$ is the radius of the 3\nobreakdash-sphere, the final placement of the player's head is given by the product of the four rotation matrices

%
%
{\footnotesize
\[
	\left(
		\begin{array}{cccc}
			 \cos \theta & 0 & 0 & -\sin \theta \\
			      0      & 1 & 0 & \phantom{-}      0      \\
			      0      & 0 & 1 & \phantom{-}      0      \\
			 \sin \theta & 0 & 0 & \phantom{-}\cos \theta
		\end{array}
	\right)
	\boldsymbol{\cdot}
	\left(
		\begin{array}{cccc}
			 1 & \phantom{-}      0      & 0 &       0      \\
			 0 & \phantom{-}\cos \theta  & 0 &  \sin \theta \\
			 0 & \phantom{-}      0      & 1 &      0      \\
			 0 & -\sin \theta            & 0 &  \cos \theta
		\end{array}
	\right)
	\quad 
\]
\[
	\quad 
	\boldsymbol{\cdot}
	\left(
		\begin{array}{cccc}
			\phantom{-}\cos \theta & 0 & 0 &  \sin \theta \\
			\phantom{-}      0      & 1 & 0 &       0      \\
			\phantom{-}      0      & 0 & 1 &       0      \\
			-\sin \theta & 0 & 0 &  \cos \theta
		\end{array}
	\right)
	\boldsymbol{\cdot}
	\left(
		\begin{array}{cccc}
			 1 &       0      & 0 & \phantom{-}      0      \\
			 0 &  \cos \theta & 0 & -\sin \theta \\
			 0 &       0      & 1 & \phantom{-}      0      \\
			 0 &  \sin \theta & 0 & \phantom{-}\cos \theta
		\end{array}
	\right)
\]
}

\noindent This product assumes the left-to-right convention in which matrices act as

\smallskip

\centerline{\textit{(row vector)}(first factor)(second factor)\ldots}

\smallskip

\noindent but may also be sensibly interpreted using the right-to-left convention

\smallskip

\centerline{\ldots(second factor)(first factor)\textit{(column vector)}}

\smallskip

\noindent Either way, the matrices for the forward, leftward, backward, and rightward motions of the player's head get applied in the opposite order from which the player makes those motions, which is somewhat counterintuitive but nevertheless correct.  The distance $d$ is typically small relative to the 3\nobreakdash-sphere's radius $R$, in which case the matrix product multiplies out to approximately
\[
	\left(
		\begin{array}{cccc}
			          1          &       \theta^2       & 0 & -\frac{\theta^{3}}{2} \\[4pt]
			      -\theta^2      &          1           & 0 & -\frac{\theta^{3}}{2} \\[4pt]
			          0          &          0           & 1 &       0               \\[4pt]
			\frac{\theta^{3}}{2} & \frac{\theta^{3}}{2} & 0 &       1
		\end{array}
	\right)
\]
The entries in the bottom row tell us that the player's virtual head ends up offset from its original position by about $\frac{1}{2}\theta^{3}$ radians in both the $x$ and $y$ directions, while the upper-left $3\times 3$ block tells us that the player's head ends up rotated by an angle of approximately $\theta^2$ radians.  While the absolute offset is only a third-order effect, the offset as a fraction of the square's width $\theta$, namely $\frac{1}{2}\theta^{3}/\theta = \frac{1}{2}\theta^{2}$ is a second-order effect, just like the rotation.

The preceding computation proves that when the player moves her head around a small square in the physical Euclidean lab, as shown in Figure~\ref{Fig:HolonomyLab}, the naive head-tracking algorithm would move her virtual head around a path in the 3\nobreakdash-sphere like the one shown in Figure~\ref{Fig:HolonomySim}(a).  In the hyperbolic case, an analogous matrix product, but this time using Lorentz matrices

%
%
{\tiny
\[
	\left(
		\begin{array}{cccc}
			 \cosh \theta & 0 & 0 & \sinh \theta \\
			      0       & 1 & 0 &       0      \\
			      0       & 0 & 1 &       0      \\
			 \sinh \theta & 0 & 0 & \cosh \theta
		\end{array}
	\right)
	\boldsymbol{\cdot}
	\left(
		\begin{array}{cccc}
			 1 & \phantom{-}      0      & 0 &  \phantom{-}     0      \\
			 0 & \phantom{-}\cosh \theta & 0 &  -\sinh \theta          \\
			 0 & \phantom{-}      0      & 1 &  \phantom{-}     0      \\
			 0 & -\sinh \theta           & 0 &  \phantom{-}\cosh \theta
		\end{array}
	\right)
	\quad 
\]
\[
	\quad 
	\boldsymbol{\cdot}
	\left(
		\begin{array}{cccc}
			\phantom{-}\cosh \theta & 0 & 0 &  -\sinh \theta          \\
			\phantom{-}      0      & 1 & 0 &  \phantom{-}     0      \\
			\phantom{-}      0      & 0 & 1 &  \phantom{-}     0      \\
			-\sinh \theta           & 0 & 0 &  \phantom{-}\cosh \theta
		\end{array}
	\right)
	\boldsymbol{\cdot}
	\left(
		\begin{array}{cccc}
			 1 &       0       & 0 &       0      \\
			 0 &  \cosh \theta & 0 & \sinh \theta \\
			 0 &       0       & 1 &       0      \\
			 0 &  \sinh \theta & 0 & \cosh \theta
		\end{array}
	\right)
\]
}

\noindent shows that the naive head-tracking algorithm would move the player's virtual head around a path in hyperbolic 3\nobreakdash-space like the one shown in Figure~\ref{Fig:HolonomySim}(b).  In both cases, when the player's head returns to its starting point in the physical Euclidean lab, her virtual head ends up slightly offset and slightly rotated relative to its original pose in the virtual curved space.  In and of itself that's not a problem, but if she keeps her hands still while moving her head, then she does have a problem:  she \textit{sees} her own hands sitting slightly offset and slightly rotated (Figure~\ref{Fig:HolonomySim}), even though she still \textit{feels} her hands sitting straight in front of her (Figure~\ref{Fig:HolonomyLab}).

This discrepancy between what the player sees and what she feels, we call \textit{body incoherence}.  It's a challenging problem that all interactive curved-space VR simulations, present and future, will face.  Section~\ref{Sec:BodyCoherenceSolution} presents a solution that works well in the case of curved-space billiards, along with some more general suggestions for other games.

\section{Headset tracking and body coherence: a solution}
\label{Sec:BodyCoherenceSolution}

If you could truly visit hyperbolic 3\nobreakdash-space or a 3\nobreakdash-sphere, would you really experience body incoherence?  No, of course not.  What you'd see and what you'd feel would remain perfectly coherent.  But as you parallel-translated your head around in a small circle, you'd feel a slight torque in your neck.  You'd need to use your neck muscles to counter that torque, to keep your head pointed straight relative to your body.  This mysterious uninvited torque would feel roughly analogous to what you feel when you rotate a spinning gyroscope about one axis and must resist a mysterious uninvited torque to prevent the gyroscope from rotating about a perpendicular axis.

Curved-space simulations, such as the curved-space billiards game, must pretend that the user's neck is providing whatever torque is necessary to keep her head aligned with her body.  Hence we must abandon the naive tracking algorithm and devise a new algorithm that's guaranteed to place the user's head and hands coherently in the curved virtual space.

\subsection{Tracking algorithm for curved-space billiards}
\label{Sec:TrackingInBilliards}

\begin{figure}[hb!]
	\centering
	\includegraphics[width=0.4\columnwidth]{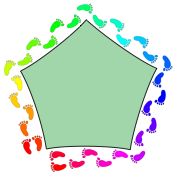}
	\hspace{0.02\linewidth}
	\includegraphics[width=0.4\columnwidth]{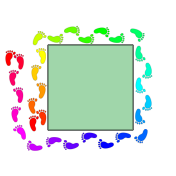}
	\caption{In order to walk one full lap around the pentagonal virtual table, the player must walk one-and-a-quarter laps around the square physical table.}
	\label{Fig:WalkingAroundTable}
\end{figure}

The curved-space billiards game incorporates a real physical table into its simulation (Figure~\ref{Fig:HyperbolicBilliardsPhoto}) to add a tactile component to the game.  You might be wondering how a physical square table in $\mathrm{E}^3$ (Figure~\ref{Fig:HyperbolicBilliardsPhoto}) could possibly be reconciled with a virtual pentagonal table in $\mathrm{H}^3$ (Figure~\ref{Fig:TableEdgeOn}), given that the physical table has only four corners while the virtual table has five, and their geometries are different.  In spite of that, amazingly enough, their perimeters are locally identical!  Both tables' perimeters consist of a sequence of 90~cm edges meeting at 90\degree{} angles. Thus the player may walk around the pentagonal virtual table, constantly touching the physical table's edges and corners with her hands as she goes around, and what she feels with her hands will always agree with she sees with her eyes. Of course, a player who walks one full lap around the pentagonal virtual table will be walking one-and-a-quarter times around the square physical table (Figure~\ref{Fig:WalkingAroundTable}), but the player will be wearing a VR headset and won't see the physical table at all, only the virtual one. As a practical matter, the physical table also gives the player a place to rest her ``bridge hand'' while setting up a shot. By combining visual and tactile sensations in this way, the experience of playing billiards on a hyperbolic billiards table becomes all the more convincing.

Analogous comments apply to a virtual triangular table in $\mathrm{S}^3$ (Figure~\ref{Fig:TriangularTable}), except that a player who walks one full lap around the triangular virtual table will have walked only three-quarters of the way around the square physical table.

\begin{figure}[t!]
	\centering
	\includegraphics[width=0.4\linewidth]{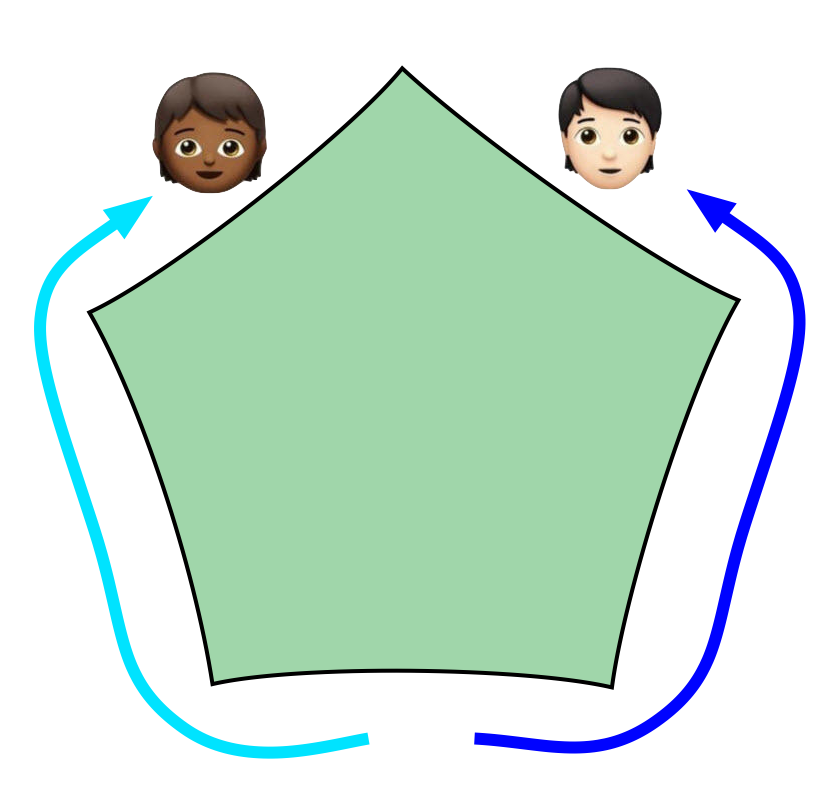}
	\hspace{0.02\linewidth}
	\includegraphics[width=0.4\linewidth]{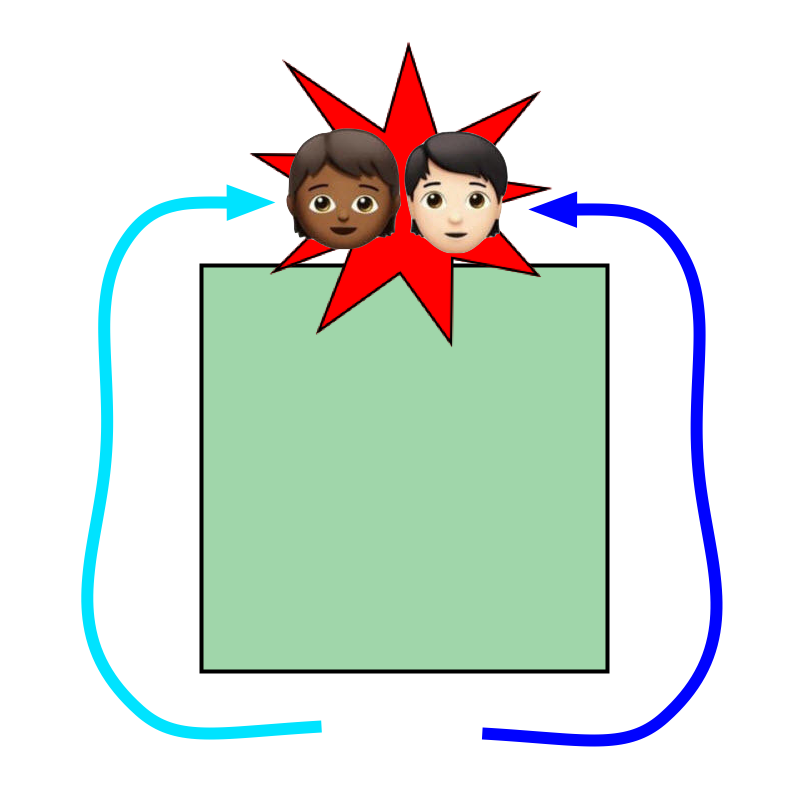}
	\caption{Even while staying a safe distance apart in the virtual space,
	  two players could accidentally collide if there were playing
	  at the same physical table.}
	\label{Fig:Collision}
\end{figure}

Two players could not share the same physical billiards table, because even while taking great care not to bump into each other in the virtual space (Figure~\ref{Fig:Collision}~left), they might nevertheless collide in the physical world (Figure~\ref{Fig:Collision}~right).  A future release version of the billiards game will, however, let two players at separate physical tables play billiards together at the same virtual table in a shared virtual space.  Each player will see the other player's face as a avatar (typically a photo) in the shared world.

\medskip

The presence of the square physical table means that we must ensure not only the coherence of the player's own body, but also the coherence between where she sees the virtual table with her eyes and where she feels the physical table with her hands.  The simplest algorithm is to track each object---the player's head and each of her two hands---relative to the nearest edge of the physical table.  This ensures that the player's head and hands stay coherent relative to the table edge, which in turn ensures that they stay coherent relative to each other as well.


\begin{figure*}[t!]
\centering

\begin{minipage}[c]{0.36\linewidth}
	\centering
	\includegraphics[width=\linewidth]{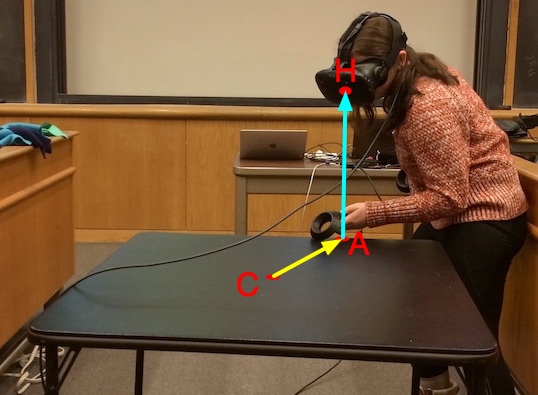}
\end{minipage}
\hspace{0.03\linewidth}	
\begin{minipage}[c]{0.36\linewidth}
	\centering
	\includegraphics[width=\linewidth]{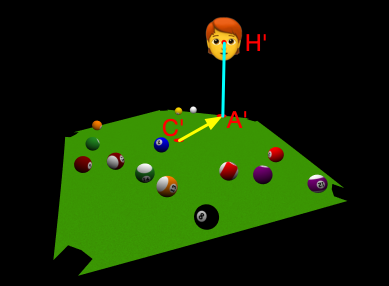}
\end{minipage}

\smallskip

\begin{minipage}[c]{0.36\linewidth} 
	\centering
	\footnotesize\textit{(a) The tracking algorithm
		expresses the player's physical head position $H$ relative 
		to a nearby point $A$ on an edge of the square physical table\ldots}
\end{minipage}
\hspace{0.03\linewidth}	
\begin{minipage}[c]{0.36\linewidth}
	\centering
	\footnotesize\textit{(b) \ldots and then
		places the player's virtual head $H'$ in an analogous position
		relative to the corresponding point $A'$
		on an edge of the pentagonal virtual table.}
\end{minipage}

\caption{The tracking algorithm as applied to billiards}
\label{Fig:Tracking}
\end{figure*}


For concreteness, the following five steps define the tracking algorithm as applied to the player's head, but it applies equally well to the player's hands.

\bigskip

\centerline{\textbf{Tracking Algorithm}}
\centerline{\textit{for billiards}}
\begin{enumerate}

\item  From the center~$C$ of the physical table, extend a horizontal ray in the direction of the player's head~$H$.  Let the point where that ray crosses the table's edge be our \textit{physical anchor point}~$A$ (Figure~\ref{Fig:Tracking}(a)).

\item  Express the player's physical head pose relative to~$A$ as the composition of
\begin{itemize}[noitemsep,topsep=0pt]
\item a rotation $R$ about~$A$ (3 degrees of freedom), and
\item a translation $T$ from $A$ to $H$.
\end{itemize}

\item Decide which edge~$E'$ of the virtual table corresponds to the edge~$E$ of the physical table that contains the anchor point~$A$.  The text below will explain how to do this.

\item Let our \textit{virtual anchor point}~$A'$ be the point that sits the same distance along $E'$ that the physical anchor point $A$ sits along $E$.  (Recall that $E$ and $E'$ have the same length.)

\item Compute the player's virtual head pose by starting at the virtual anchor point~$A'$ and applying the rotation $R$ and translation $T$ from Step~2, but re-interpreted here as rigid motions of the curved virtual space (Figure~\ref{Fig:Tracking}(b)).

\end{enumerate}

The question remains, in Step~3 how do we decide which edge~$E'$ of the virtual table corresponds to edge~$E$ of the physical table?  The correspondence between the edges of the square physical table and the edges of the virtual triangular, square or pentagonal table is not fixed, but is something that must be traced around in real time, as the player walks around the table (Figure~\ref{Fig:WalkingAroundTable}). For example, if a player keeps walking around the physical table---always in the same direction, never turning back---she'll see the following edges in the following order:

{\footnotesize
\[
	\begin{array}{ll}
		\text{square physical table:}     & \text{0 1 2 3 0 1 2 3 0 1 2 3 0 1 2 \ldots} \\
		\text{pentagonal virtual table}:  & \text{0 1 2 3 4 0 1 2 3 4 0 1 2 3 4 \ldots}
	\end{array}
\]
}

This matching of an edge of the physical table to an edge of the virtual table is the only thing that depends on the player's history.  Once that matching is known, the headset's pose in the Euclidean physical space uniquely determines its pose in the curved virtual space, as specified in Steps 1, 2, 4 and 5 above, with no additional history dependence.

We apply the same tracking algorithm to the player's left and right hands, just as we applied it to her head, but with one difference:  The correspondence between the edges of the physical table and the edges of the virtual table gets decided only once, based on the player's head position, and then that same correspondence gets re-used in Step~3 of the algorithm when computing the hand poses.  During normal play this ad~hoc rule makes absolutely no difference.  The only time it would have an effect would be if a mischievous player kept her head on one side of the table while running one of her hands all the way around the physical table's perimeter.  Without the ad~hoc rule, when she felt her physical hand back in front of her, her eyes would see her virtual hand sitting near the wrong edge of the virtual table.  With the ad~hoc rule, when her hand passes the far side of the table, the virtual image of that hand suffers a discontinuous jump, to ensure that the virtual hand remains coherent with the physical hand as it completes its lap around the table.

\subsection{General principles}
\label{Sec:TrackingInGeneral}

Looking to the future, all designers of interactive curved-space VR games will face the body coherence problem.  In broad outline, our solution applies to all games, in all spaces (even non\nobreakdash-isotropic spaces):

\bigskip

\centerline{\textbf{Tracking Algorithm}}
\centerline{\textit{in general}}
\begin{enumerate}

\item Use the current position of the player's head and hands to define a physical anchor point~$A$.  The definition could be as simple as letting the player's head be the anchor point, or it could depend on the player's hands and/or other objects in the scene.

\item Express the player's physical head and hand poses relative to the physical anchor point~$A$.

\item Let $\Delta A$ be the motion that takes the physical anchor point~$A_{prev}$ from the previous frame to the physical anchor point~$A$ of the current frame, expressed in local coordinates appropriate to the anchor's definition.

\item Apply that same incremental motion $\Delta A$ to the previous virtual anchor point~$A'_{prev}$ to get the current virtual anchor point~$A'$.  Here we're working in the virtual anchor's local coordinates, and the increment $\Delta A$ is expressed as a rigid motion of the curved virtual space.

\item Compute the player's virtual head and hand poses relative to the virtual anchor point~$A'$.  The player's virtual head and hands should bear the same relationship to the virtual anchor point~$A'$ that her physical head and hands bear to the physical anchor point~$A$.

\end{enumerate}

For simple curved-space games, it may be enough to define the anchor point~$A$ to coincide with the player's head, in which case the above algorithm lets the player's head move freely in the curved space while ensuring that her hands follow along coherently.  For more complex games (like the billiards game, which requires not only internal body coherence, but also coherence with a physical table), other considerations come into play.  While a particular app's ideal solution will depend on its particular needs, some general principles apply:

\begin{itemize}

\item\textit{Ensure visual stability.}  In a scene with no fixed reference points, it's better to let the player's virtual head move freely and place the player's virtual hands to stay consistent with it, rather than to let the hands move freely and place the head to stay consistent.  The reason for this is that artificially tweaking the player's head would slightly shift her view of the whole scene.  Hand motion should, if possible, have no effect on how the player sees the scene.  But if head motion causes a slight change in the position of the player's virtual hands, that would be undesirable but ultimately acceptable.  This asymmetry is due to the primacy of the human visual system:  a person's sense of ``where she is in the world'' is tied most strongly to what she sees;  her hands and feet are then perceived as being 50--150\,cm below that primary viewpoint.

\item\textit{Keep the player away from the walls.}  If the player in the curved-space billiards game weren't anchored to the table, then if she parallel-translated herself around a small circle several times, she'd see the virtual table orbit around her, eventually passing behind her back.  At that point, if she turned around to face the table and take a shot, she'd risk walking into a wall of the physical lab (or, more likely, she'd see the ``chaperone'' bounds that the VR system inserts into the scene to prevent players from walking into walls).  To avoid such a scenario, all curved-space VR games should, if possible, keep their primary virtual content centered in the middle of the physical lab, and keep the player's position coherent with that primary content.

\end{itemize}

The most challenging curved-space VR apps to design may be those whose game content is inherently large-scale, such as a potential VR version of HyperRogue \citep{KopczynskiCelinska2019}.  Even games whose content may be only a few meters across, such as the author's 3D mazes in any of several multi-connected spaces, nevertheless appear infinite to the player, because the player sees such a maze as an infinite periodic structure, beckoning her to wander beyond the bounds of the fundamental domain (\url{http://www.geometrygames.org/TorusGames}).  How might we let the player travel longer distances than her real-world room would permit?  One solution would be to give her the ability to press a button and fly through the scene, but that approach risks motion sickness, due to the player's eyes seeing accelerations that her inner ear doesn't feel.  A safer solution---one used in some flat-space VR apps---would be \textit{teleportation} \citep{BoletsisCedergren2019}:  after selecting a destination and pressing a button, the player ``fades out'' from her current location, briefly sees total darkness, and then ``fades back in'' at her selected destination.  More sophisticated flat-space techniques, such as \textit{saccadic redirection} \citep{SunEtAl2018}, could be applied in curved space as well.

\section{Computer graphics in curved space}
\label{Sec:CurvedSpaceGraphics}

Our approach to curved-space graphics uses the same sequence of transformations that flat-space graphics uses:
\begin{align*}
 \text{Model Space} & \longrightarrow \text{World Space} \\
                    & \longrightarrow \text{Eye Space} \\
                    & \longrightarrow \text{Projection Space}
\end{align*}
An elementary exposition of computer graphics in $\mathrm{S}^3$, $\mathrm{E}^3$ and $\mathrm{H}^3$ appears in \citep{Weeks2002}.  The following subsections summarize the essentials of that article, but with a new approach to radians in the Euclidean case, a comment on units in VR, a new way to visualize the projection transformation, and an algorithm for drawing the ``back hemisphere'' in the spherical case.

\textit{Note:}  While recent ray-tracing \citep{KopczynskiCelinska2020, NovelloEtAl2020b} and ray-marching \citep{CoulonEtAl2020c} work has produced beautiful curved-space animations, as of early 2021 a traditional rasterization approach remains more efficient on consumer-level VR hardware, so we'll focus on that.

\subsection{Euclidean radians}
\label{Subsec:Radians}

A game like curved-space billiards typically uses \textit{exactly} the same computer code for all three spaces ($\mathrm{S}^3$, $\mathrm{E}^3$ and $\mathrm{H}^3$), with no need to split into separate cases.  To make this possible, we need to change how we think about---and implement---Euclidean translations, to make them consistent with how we handle translations in the spherical and hyperbolic cases.

In the spherical case, the model, world, and eye spaces are
all 3\nobreakdash-spheres of some desired radius $\rho$, sitting in Euclidean 4\nobreakdash-space (recall Figure~\ref{Fig:Radii}(left)).  The model and view matrices are elements of the rotation group $\mathrm{O}(4)$.

In the hyperbolic case, the model, world, and eye spaces are all hyperbolic 3\nobreakdash-spaces of some desired radius $\rho$, sitting in the Minkowski space (recall Figure~\ref{Fig:Radii}(right)).  The model and view matrices are elements of the Lorentz group $\mathrm{O}(3,1)$.

Surprisingly, it's the Euclidean case that requires the greatest care.
A standard computer graphics trick is to take the coordinates $(x,y,z)$ and append a ``1'' as a fourth coordinate, so we can package up a rotation $R$ and a translation $(\Delta x~\Delta y~\Delta z)$ as a single $4\times 4$ matrix
\[
	\left(x\;y\;z\;1\right)
	\left(
		\begin{array}{cccc}
			R_{00} & R_{01} & R_{02} & 0 \\
			R_{10} & R_{11} & R_{12} & 0 \\
			R_{20} & R_{21} & R_{22} & 0 \\
			\Delta x & \Delta y & \Delta z & 1
		\end{array}
	\right)
\]
When you do the matrix multiplication, the ``1'' multiplies the $\Delta x~\Delta y~\Delta z$ and they get added in.
The awkward question here is:
What are the units on $\Delta x$, $\Delta y$, and $\Delta z$?
If we're measuring $x$, $y$, and $z$ in meters, are $\Delta x$, $\Delta y$, and $\Delta z$ also in meters?
Yet the rotational components $R_{ij}$ are dimensionless.
Having a transformation matrix with some dimensionless entries and other entries in meters gives us a hint that our understanding of that matrix is less than complete.
The situation becomes much clearer if instead of representing Euclidean 3\nobreakdash-space as a hyperplane at height $w = 1$, we represent it as a hyperplane at height $w =\rho$, and call $\rho$ the ``radius'' of this Euclidean space (Figure~\ref{Fig:Radii}(center)).
So now we can rewrite our matrix product as
\[
	\left(x\;y\;z\;\rho\right)
	\left(
		\begin{array}{cccc}
			R_{00} & R_{01} & R_{02} & 0 \\
			R_{10} & R_{11} & R_{12} & 0 \\
			R_{20} & R_{21} & R_{22} & 0 \\
			\Delta x/\rho & \Delta y/\rho & \Delta z/\rho & 1
		\end{array}
	\right)
\]
and it's abundantly clear that $\rho$ is in meters, just like $x$, $y$, and $z$, and $\Delta x/\rho$, $\Delta y/\rho$, and $\Delta z/\rho$ are dimensionless, just like the $R_{ij}$.
The best way to think of this is that $\Delta x/\rho$, $\Delta y/\rho$, and $\Delta z/\rho$ are translation distances in \textit{Euclidean radians}.

The big practical advantage to using Euclidean radians is that, by having a purely dimensionless transformation matrix acting on a position vector $(x, y, z, \rho)$ in pure meters, we make the Euclidean case consistent with the spherical and hyperbolic cases, which lets us use the same computer code for all three spaces.

\subsection{VR needs explicit units}
\label{Subsec:Units}

When writing traditional non-VR animations, most mathematicians (including the author) treated distances as dimensionless quantities, in effect measuring all distances in spherical, hyperbolic, or Euclidean radians.  And that was fine:  in a non-VR maze in a 3-torus or a non-VR flight through a curved space, explicit units aren't needed.

In VR, by contrast, units are essential.  The player's two eyes and two hands are immersed in the simulated world, so the distances the player sees with her eyes must agree with the distances she feels with her hands.  The player perceives her own physical body as part of the scene.

For an acceptable VR simulation, we must know (1) the space's radius of curvature in meters and (2) the size of each object in meters.  To see why, consider what happens when a player in a hyperbolic billiards game changes the radius of curvature of the space.  If she increases the radius of curvature, the space as a whole will scale up, and any geometrical structures whose size is tied to the radius of curvature will also scale up.  For example, the billiards table, which we've defined to be a right-angled regular pentagon, will get larger.  But the player's body, whose size in meters is fixed, will not get larger.  Nor will the billiard balls, whose diameter is fixed at 57\,mm, nor the cue stick, whose length is fixed at 1\,m.  The final result is that when the player increases the space's radius of curvature, she'll find herself playing billiards on a larger right-angled pentagon, but with the same familiar billiard balls and cue stick.

\subsection{Visualizing the projection transformation}
\label{Subsec:ProjTrans}

As part of the standard graphics pipeline, a \textit{vertex shader} computes and reports each vertex's position in homogeneous coordinates $(x,y,z,w)$.  The Graphics Processing Unit (GPU) then divides through by the $w$\nobreakdash-coordinate
\[
	(x,y,z,w) \mapsto \left(\frac{x}{w}, \frac{y}{w}, \frac{z}{w}, 1\right)
\]
to map the desired \textit{view frustum} onto a rectangular box in the $w = 1$ hyperplane.

Even though the GPU divides through by $w$, I personally find it easier to think in terms of dividing through by $z$ instead.  Dividing through by $z$ makes the projection's geometrical effect much clearer.  Moreover, essentially all vertex shaders apply a 90\degree{} rotation in the $zw$\nobreakdash-plane \textit{before} the GPU divides through by $w$, to ensure that the GPU's hardwired division-by-$w$ reproduces the same effect as our geometrically clear division-by-$z$.  Conceptually, the projection formula for all three spaces ($\mathrm{S}^3$, $\mathrm{E}^3$ and $\mathrm{H}^3$) becomes
\[
	(x,y,z,w) \mapsto \left(\frac{x}{z}, \frac{y}{z}, 1, \frac{w}{z}\right)
\]

\begin{figure}[ht!]
	\centering
	\includegraphics[width=0.68\linewidth]{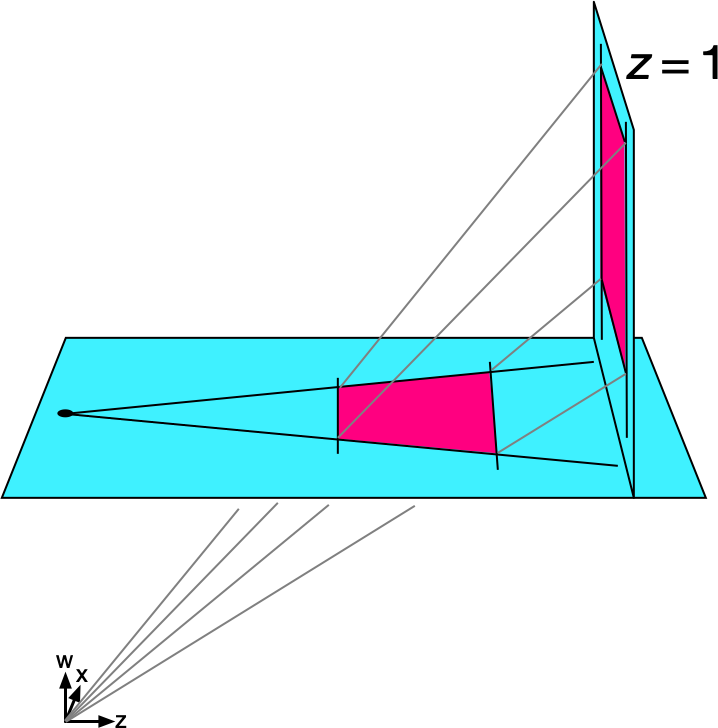}
	\caption{In Euclidean graphics, dividing by \textit{z} maps the view frustum to a rectangular box (\textit{y}~coordinate suppressed).}
	\label{Fig:ProjectionEuc}
\end{figure}

In the Euclidean case, the division-by-$z$ takes the view frustum to a rectangular box, as desired (Figure~\ref{Fig:ProjectionEuc}).  It's perfectly acceptable to push the frustum's far wall off to infinity, thus mapping an infinite portion of a pyramid onto a finite rectangular box.   By contrast, if the frustum's near wall gets too close to the observer, it will push the rectangular box's top face upwards towards infinity, which quickly degrades the numerical precision of the results as the floating-point arithmetic's finite precision gets spread out over an ever taller box.

\begin{figure}[hb!]
	\centering
	\includegraphics[width=0.68\linewidth]{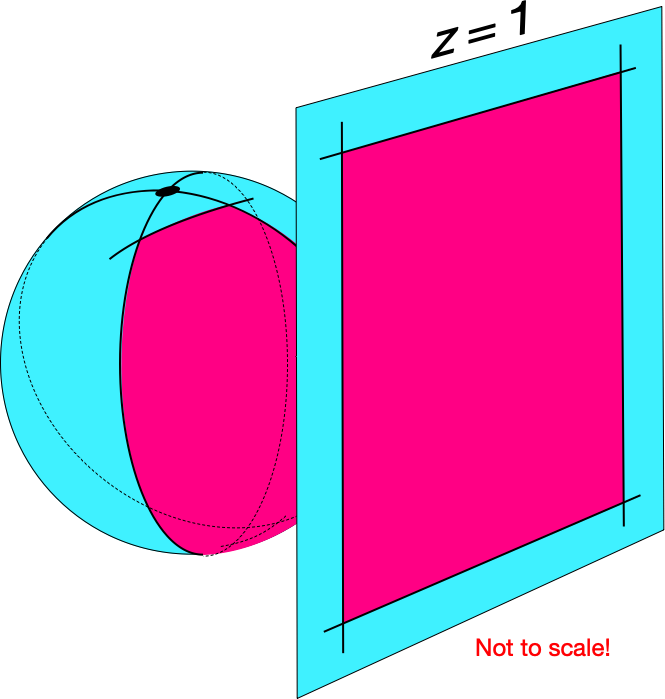}
	\caption{In spherical graphics, dividing by \textit{z} maps the view banana 
			to a rectangular box (\textit{y}~coordinate suppressed).}
	\label{Fig:ProjectionSph}
\end{figure}

In the spherical case, that same division-by-$z$ takes not a frustum, but rather a solid I like to call the \textit{view banana} (Figure~\ref{Fig:ProjectionSph}), to a rectangular box.  Like a real banana, the view banana has flat sides and tapers down at both ends.

\begin{figure}[ht!]
	\centering
	\includegraphics[width=0.76\linewidth]{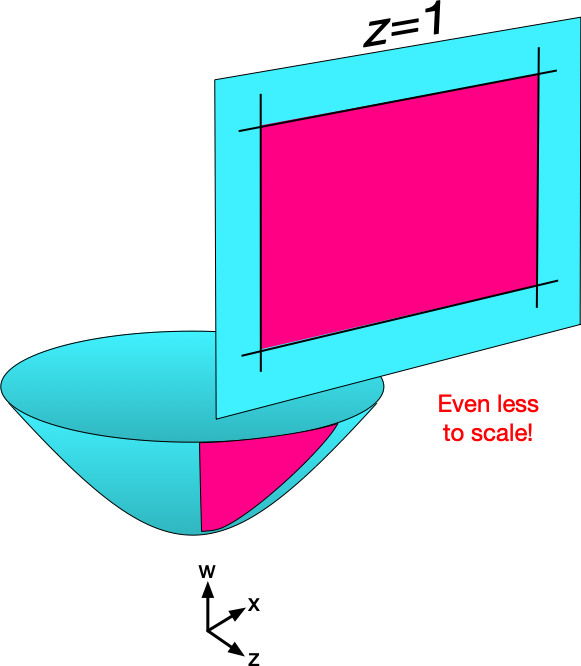}
	\caption{In hyperbolic graphics, dividing by \textit{z} maps the view frustum
			to a rectangular box (\textit{y}~coordinate suppressed).}
	\label{Fig:ProjectionHyp}
\end{figure}

In the hyperbolic case (Figure~\ref{Fig:ProjectionHyp}), we again map a frustum-like solid onto a rectangular box, and again it's perfectly acceptable to push the frustum's far wall off to infinity.

\subsection{The 3-sphere}
\label{Subsec:BackHemi}

Rendering images in the 3\nobreakdash-sphere is a two-part process.  To see why, first recall that in Euclidean graphics, if we let the near clipping distance go to zero, the short edge of the red trapezoid in Figure~\ref{Fig:ProjectionEuc} would approach the observer (the black dot), which would send the top edge of its projected image (the top edge of the red rectangle in the vertical plane) upwards to infinity.  To avoid that problem, in Euclidean graphics we always set the near clipping distance to some $\varepsilon > 0$.

In spherical graphics, we face that same problem at both the top and the bottom of the red view banana on the sphere in Figure~\ref{Fig:ProjectionSph}.  To avoid trouble, we must set the near clipping distance to some $\varepsilon > 0$ to keep the top edge of its projected image (the top edge of the red rectangle in the vertical plane) from going upwards to infinity, and we must also set the far clipping distance to $\pi - \varepsilon$ to keep the bottom edge of the projected image from going downwards to infinity.

One immediate consequence of this approach is that an $\varepsilon$\nobreakdash-neighborhood of the observer's antipodal point is always excluded from the rendered image.  This is not surprising, given that a tiny fleck of dust sitting at the antipodal point would fill the observer's entire sky!  While ray-tracing methods would have no problem rendering the antipodal point, our mesh-based methods always omit the antipodal point and some tiny neighborhood surrounding it.

Our mesh-based methods can, however, easily see through the antipodal point and render content that sits on the ``back side'' of the 3\nobreakdash-sphere.  That is, we may render content whose line-of-sight distance from the observer falls in the range $[{\pi + \varepsilon}, {2\pi - \varepsilon}]$.  Because the observer's lines-of-sight all reconverge at the antipodal point, what the observer sees at distances $[{\pi + \varepsilon}, {2\pi - \varepsilon}]$ is precisely the same as what her antipodal twin would see at distances $[\varepsilon, {\pi - \varepsilon}]$.  In other words, to render the back hemisphere, we may simply apply our usual methods to render what the antipodal twin sees at distances $[\varepsilon, {\pi - \varepsilon}]$.

\textit{Technical detail:}  The most efficient way to implement this on a Tile-Based Deferred Render is to encode commands taking the 3\nobreakdash-sphere's back hemisphere into the back half of the clipping box ($\frac{1}{2} \leq z \leq 1$) and the front hemisphere into the front half of the clipping box ($0 \leq z \leq \frac{1}{2}$), and then let the renderer process of the whole scene at once.  That way back-hemisphere objects that are eclipsed by front-hemisphere objects will never get rendered at all.

If the observer's head were perfectly transparent, would we also have to render objects that she sees at distances in the range $[2\pi + \varepsilon, 3\pi - \varepsilon]$?  No, this isn't necessary, because these would be the same objects that the observer sees at distances $[\varepsilon, \pi - \varepsilon]$, and they would fill exactly the same portions of observer's sky.

\section{Native-inhabitant view vs. tourist view}
\label{Sec:NativeVsTouristViews}

\begin{figure}[b!]
\centering
\includegraphics[width=0.64\linewidth]{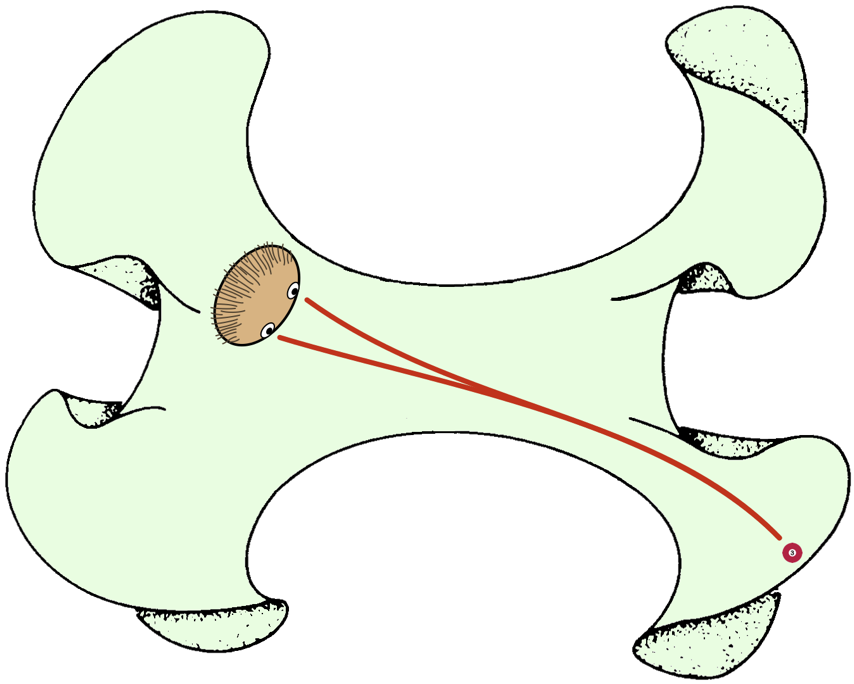}
\caption{In hyperbolic space, an observer must look slightly cross-eyed to focus on a distant object.}
\label{Fig:ParallaxHyp}
\end{figure}

To focus on a distant object in hyperbolic 3\nobreakdash-space, an observer must look slightly cross-eyed, to compensate for the space's inherent geodesic divergence (Figure~\ref{Fig:ParallaxHyp}).  For native inhabitants of hyperbolic space, this is fine, because when they were babies they learned that a certain positive \textit{vergence angle} means that the object they're looking at is far away.  But a Euclidean-born tourist who visits hyperbolic space would misinterpret that positive vergence angle as meaning that the object is close.  In fact the tourist would see all of hyperbolic space crammed into a small ball, of radius only a meter or two in the case of the hyperbolic billiards game.

\begin{figure}[b!]
\centering
\includegraphics[width=0.50\linewidth]{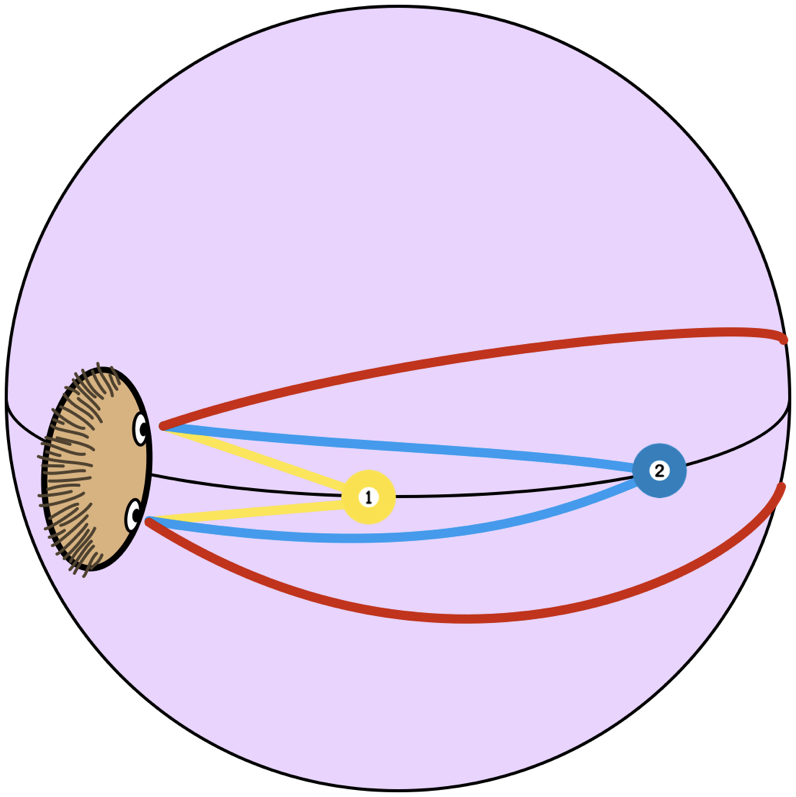}
\caption{In a 3-sphere, an observer must look ``walleyed'' to focus on objects more than 90\degree{} away.}
\label{Fig:ParallaxSph}
\end{figure}

In a 3\nobreakdash-sphere, by contrast, an observer must look less cross-eyed than in Euclidean space, to compensate for the space's inherent geodesic convergence (Figure~\ref{Fig:ParallaxSph}).  When an object is exactly 90\degree{} away, like the blue (\#2) billiard ball in Figure~\ref{Fig:ParallaxSph}, the observer doesn't need to look crosseyed at all.  For those of us who grew up in Euclidean space, our binocular vision makes that blue (\#2) ball seem infinitely far away.  If a ball is more than 90\degree{} away, like the red (\#3) ball in Figure~\ref{Fig:ParallaxSph}, the observer must look slightly ``walleyed''---with a negative vergence angle---to focus on it.

When designing a virtual reality simulation like the curved-space billiards game, the developer must decide whether to show the space the way the native sees it, to show it the way the Euclidean-born tourist sees it, or to offer the user the choice.  For me, the tourist's view is just too weird:  as you walk around hyperbolic 3\nobreakdash-space, the entire contents of the universe seem to move along with you, all trapped inside that finite ball.  Spherical space is even worse, because you have to look walleyed to focus on distant objects.  The human visual system can do this to some extent, but I personally find it uncomfortable and vaguely distressing.  To avoid such discomfort, I recommend offering the native-inhabitant view only.

To simulate a rigorously correct native-inhabitant view, we must trick the user's eyes and brain into perceiving each object's true hyperbolic or spherical distance.  This is most easily accomplished by inserting some extra steps into the graphics pipeline:
\begin{align*}
 \text{Model Space} & \longrightarrow \text{World Space} \\
                    & \longrightarrow \text{\textit{Nose Space}} \\
                    & \longrightarrow \text{\textit{Euclidean Tangent Space}} \\
                    & \longrightarrow \text{\textit{Eye Space}} \\
                    & \longrightarrow \text{Projection Space}
\end{align*}
Rather than mapping the scene's contents from the world space directly into the space of either eye, we instead map them into a space centered at the bridge of the player's nose.  From there we map them onto the nose's Euclidean tangent space in such a way that the distance from each object to the player's nose is preserved.  In cartography this is called an \textit{azimuthal equidistant map}.  We then apply a Euclidean translation of the tangent space to map the scene's contents into the space of the player's left or right eye, and finally we apply the standard Euclidean projection.

\begin{figure}[b!]
\centering
\includegraphics[width=0.5\linewidth]{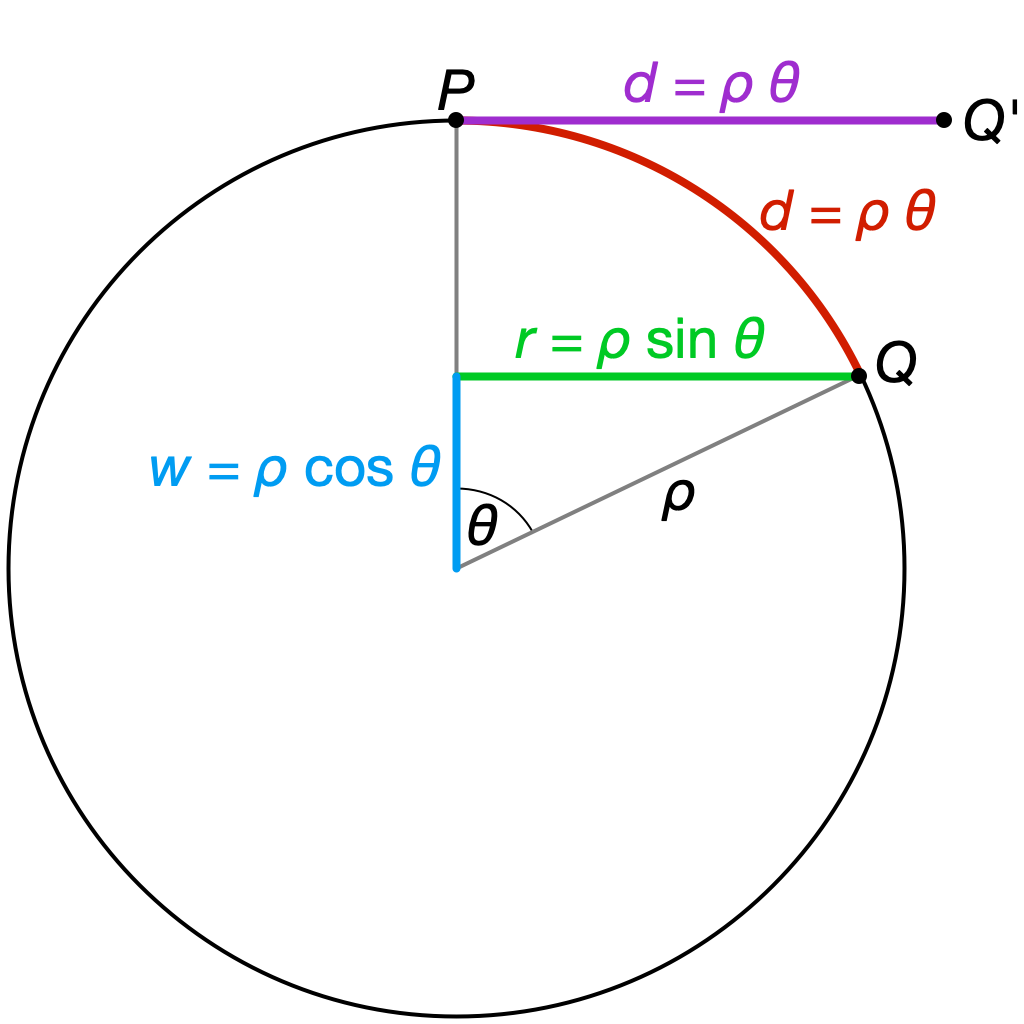}
\caption{\textit{Azimuthal equidistant map in S\textsuperscript{3}:}  The point \textit{Q}$'$ sits the same distance from \textit{P}, in the same direction, as \textit{Q} does, but \textit{Q}$'$ sits in the tangent space while \textit{Q} sits on the 3-sphere itself.}
\label{Fig:AzimuthalEquidistantMapSph}
\end{figure}

\begin{figure}[b!]
\centering
\includegraphics[width=0.5\linewidth]{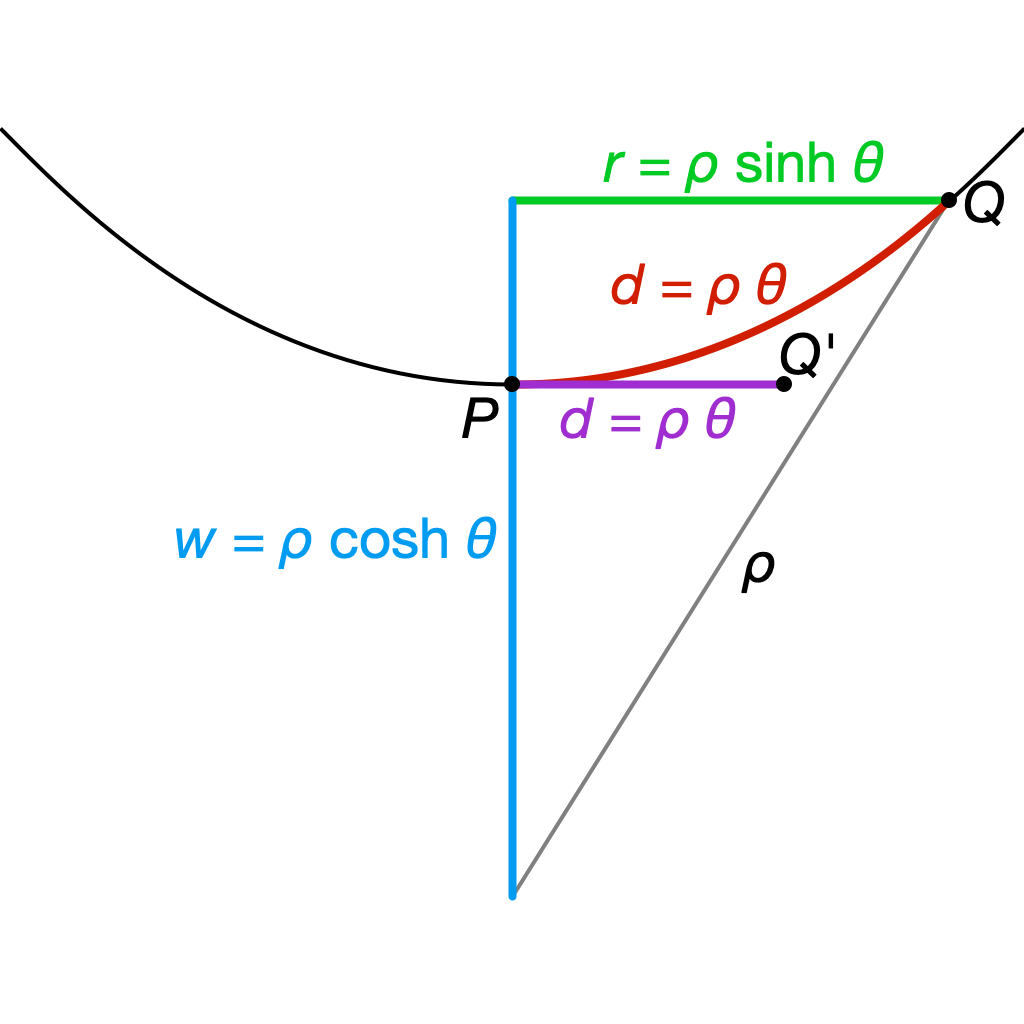}
\caption{\textit{Azimuthal equidistant map in H\textsuperscript{3}:}  This figure is essentially the same as Figure~\ref{Fig:AzimuthalEquidistantMapSph}, but drawn in Minkowski space instead of in Euclidean space.  To our Euclidean eyes, the distance along the arc from \textit{P} to \textit{Q} looks longer than the distance from \textit{P} to \textit{Q}$'$ in the tangent space, but when measured using the Lorentz metric the two distances are the same.}
\label{Fig:AzimuthalEquidistantMapHyp}
\end{figure}

While the azimuthal equidistant map cannot be realized as a matrix multiplication, it can be realized with a simple trigonometric computation, which puts only a small extra burden on the vertex shader.  For example, in the spherical case, say an observer at $P = (0,0,0,\rho)$ is viewing some point of interest $Q = (x,y,z,w)$ on a 3\nobreakdash-sphere of radius $\rho$ meters (Figure~\ref{Fig:AzimuthalEquidistantMapSph}).  Let $\theta$ be the distance in radians from $P$ to $Q$, and let $d$ be that same distance in meters (measured along the 3\nobreakdash-sphere itself, \textit{not} in the 4\nobreakdash-dimensional ball that it bounds).  We want to find a point $Q'$ in the tangent space that sits in the same direction from $P$ as $Q$ does, and also sits $d$ meters away (but in the tangent space, not on the 3\nobreakdash-sphere).

\medskip

\noindent\textbf{Proposition (for S\textsuperscript{3}).}  The desired point $Q'$, as defined above, is given by the formula
\[
    Q' = \left(\frac{\theta}{\sin\theta}\,x,\,\frac{\theta}{\sin\theta}\,y,\,\frac{\theta}{\sin\theta}\,z,\,\rho\right).
\]

\smallskip

\noindent\textit{Proof.}  Let $v = (x,y,z)$ be the ``horizontal part of $Q$'', and let $r = |v|$ be its length.  Figure~\ref{Fig:AzimuthalEquidistantMapSph} shows that $d$ = $\rho \theta$, $w = \rho \cos\theta$, and $r = \rho \sin\theta$.  Let $v'$ be the horizontal vector connecting $P$ to $Q'$.  The definition of the azimuthal equidistant map requires that $v'$ be parallel to $v$, but have length $d$ instead of length $r$.  Hence we may compute $v'$ as a scalar multiple of $v$

{\small
\[
    v' = \frac{d}{r} v
       = \frac{\rho \theta}{\rho \sin\theta} v
       = \frac{\theta}{\sin\theta} v
       = \left(\frac{\theta}{\sin\theta}\,x,\,\frac{\theta}{\sin\theta}\,y,\,\frac{\theta}{\sin\theta}\,z\right)
\]
}

\noindent Setting

{\small
\[
    Q' = P + v' = (0,0,0,\rho) + \left(\frac{\theta}{\sin\theta}\,x,\,\frac{\theta}{\sin\theta}\,y,\,\frac{\theta}{\sin\theta}\,z,\,0\right)
\]
}

\noindent gives the required formula. $\blacksquare$

\medskip

To compute $\theta$, recall that $w = \rho \cos\theta$, where $w$ and $\rho$ are both already known, so our shader code becomes

\begin{samepage}
{\small
\begin{verbatim}
  CosineTheta = clamp(w/rho, -1, +1)
  Theta = acos(CosineTheta)
  SineTheta = sin(Theta)
  if SineTheta > 0.0001
      Factor = Theta / SinTheta
  else
      Factor = 1  // correct near north pole,
                  // but not near south pole
  Qprime
    = (Factor * x, Factor * y, Factor * z, rho) 
\end{verbatim}
}
\end{samepage}

\noindent \textit{Technical note:}  When rendering the back hemisphere we must set $\mathrm{Factor} = (\pi + \theta)/\sin\theta$ in the shader code above, so that back-hemisphere objects get drawn in the azimuthal equidistant map at distances in the desired range $[{\pi + \varepsilon}, {2\pi - \varepsilon}]$.

\smallskip

The computations for the hyperbolic case (Figure~\ref{Fig:AzimuthalEquidistantMapHyp}) are essentially the same, but with \textit{sinh} and \textit{cosh} instead of \textit{sin} and \textit{cos}:

\medskip

\noindent\textbf{Proposition (for H\textsuperscript{3}).}  In the hyperbolic case, the desired point $Q'$ is given by the formula
\[
    Q' = \left(\frac{\theta}{\sinh\theta}\,x,\,\frac{\theta}{\sinh\theta}\,y,\,\frac{\theta}{\sinh\theta}\,z,\,\rho\right).
\]

\smallskip

\noindent\textit{Proof.}  The hyperbolic proof follows the spherical proof line by line, but with the geometry shown in Figure~\ref{Fig:AzimuthalEquidistantMapHyp}. $\blacksquare$

\medskip

The shader code for the hyperbolic case is

\begin{samepage}
{\small
\begin{verbatim}
  CoshTheta = clamp(w/rho, +1, +infinity)
  Theta = acosh(CoshTheta)
  SinhTheta = sinh(Theta)
  if SinhTheta > 0.0001
      Factor = Theta / SinhTheta
  else
      Factor = 1  // correct near north pole
  Qprime
    = (Factor * x, Factor * y, Factor * z, rho) 
\end{verbatim}
}
\end{samepage}

\section{Mental models of space}
\label{Sec:MentalModels}

The algorithm presented in Section~\ref{Sec:NativeVsTouristViews} lets the observer use her Euclidean binocular vision to see every point in the space at its correct hyperbolic or spherical distance.  This eliminates the illusion that all of hyperbolic space sits in a small finite ball, and also eliminates the need for the observer to look walleyed in spherical space.  However, even though the observer perceives all objects at their true hyperbolic or spherical distances, there's still the rich and interesting question of how the observer's brain integrates those distances into a mental model of the space \citep{Weeks2020b}.  The article \citep{Weeks2020b} illustrates this question with the following example:

\begin{quotation}

\noindent\ldots imagine a cue stick sitting 1~meter in front of your face in Euclidean space (Figure~\ref{Fig:CueStick}).  Point $C$ (the stick's midpoint) is closest to you, points $B$ and $D$ are slightly further away, and points $A$ and $E$ are further still.  If you repeat this experiment in spherical space, you'll observe almost the same thing\dots except that distances increase a little more slowly as you shift your attention from point $C$ to $D$ and thence to $E$.  And, by contrast, the distances increase a little more rapidly in hyperbolic space.

\begin{figure}[ht!]
	\centering
	\begin{minipage}[b]{0.75\columnwidth} 
		\centerline{\includegraphics[width=\linewidth]{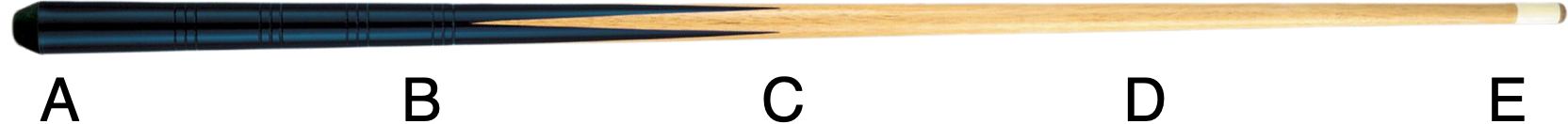}}
		\caption{The cue stick's midpoint $C$ is closest to you.
		How fast the distance increases as you shift your attention
		from point $C$ to point $D$ and thence to $E$
		depends on the curvature of the ambient space.}
		\label{Fig:CueStick}
	\end{minipage}
\end{figure}

So when you put on the VR headset and play hyperbolic billiards, even though the app is showing you the correct hyperbolic distance to every point on that cue stick, your brain chooses to misinterpret that data as a curved stick in a flat space (Figure~\ref{Fig:MentalModels}(b)), rather than as a straight stick in curved space (Figure~\ref{Fig:MentalModels}(a)).

\begin{figure}[ht!]
	\centering
	\begin{minipage}[b]{0.75\columnwidth} 

		\centerline{\includegraphics[width=\linewidth]{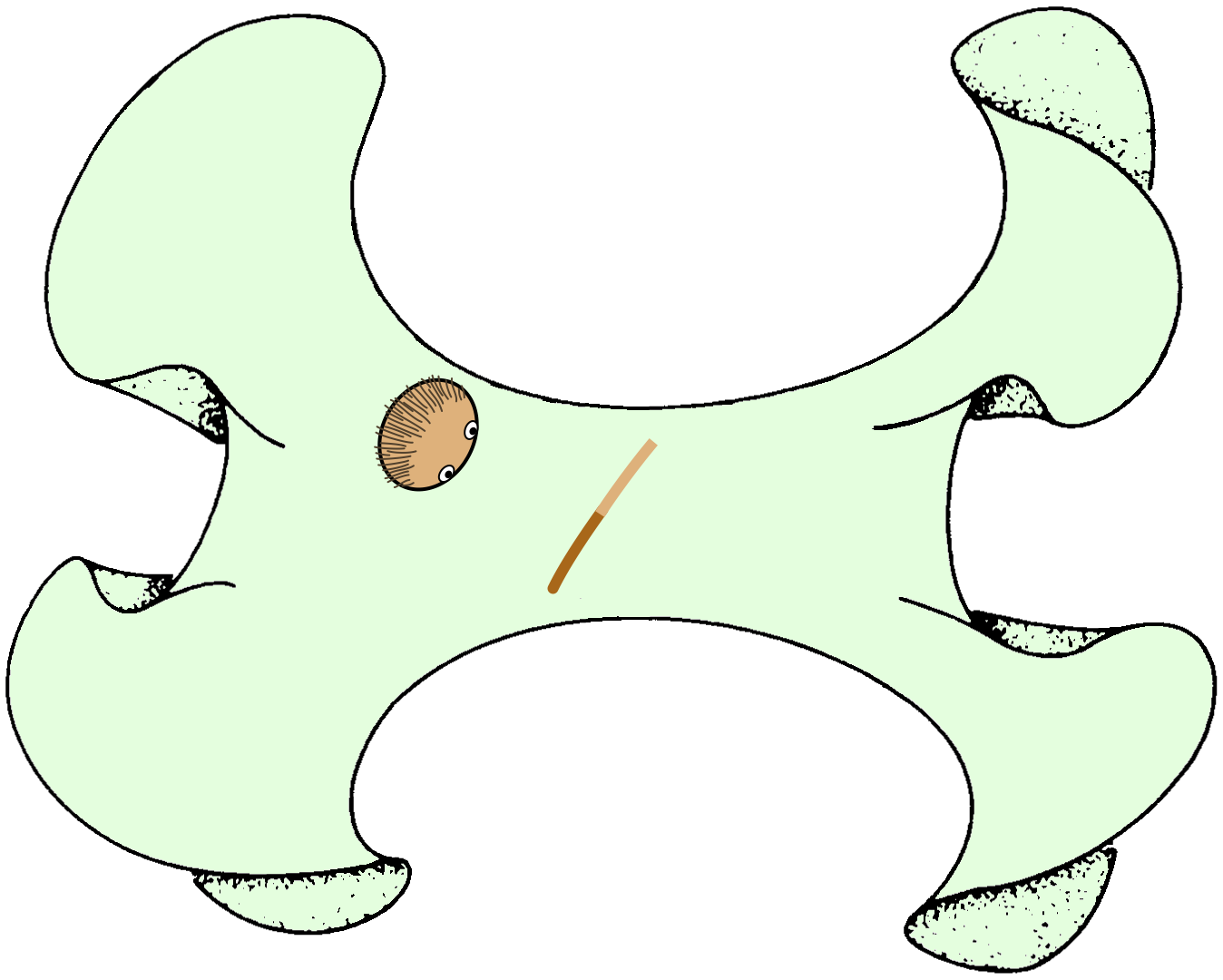}}
		{
			\footnotesize 
			\centerline{\textit{(a) A native inhabitant of hyperbolic 3\nobreakdash-space}}
			\centerline{\textit{perceives a straight cue stick in a curved space.}}
		}
		
		\vspace{0.05\columnwidth}

		\centerline{\includegraphics[width=\linewidth]{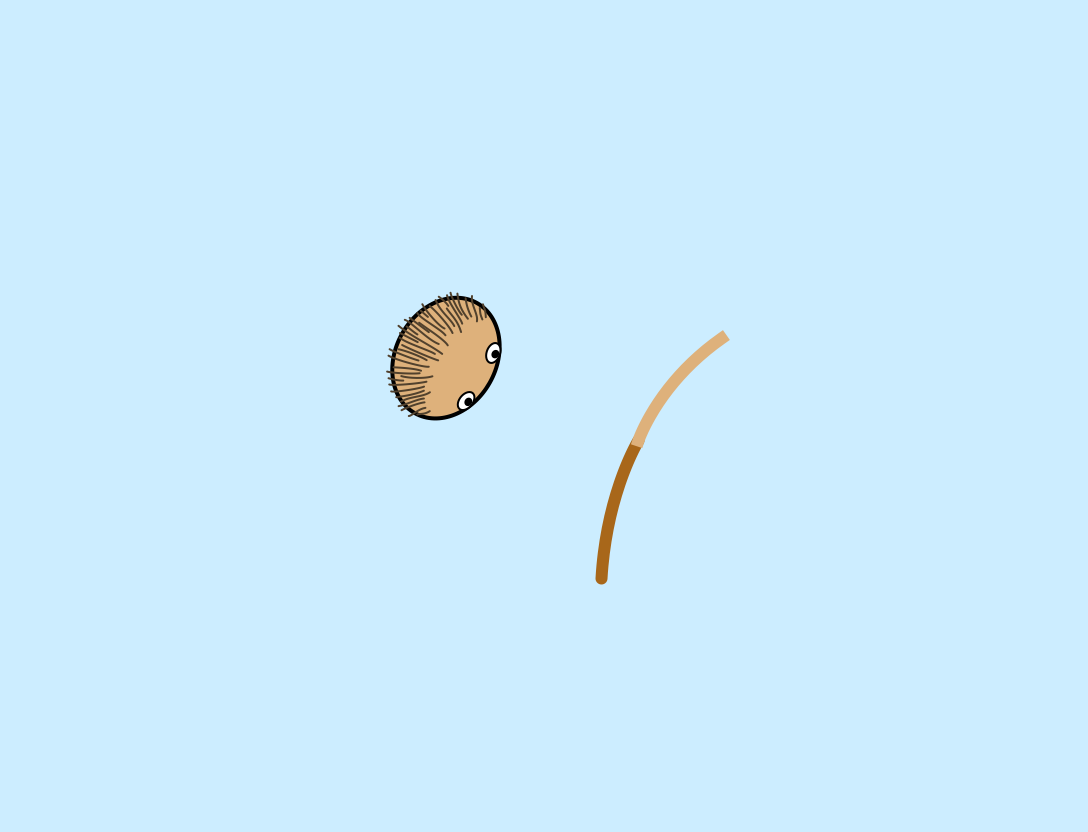}}
		{
			\footnotesize 
			\centerline{\textit{(b) A Euclidean-born tourist incorrectly}}
			\centerline{\textit{perceives a curved cue stick in a flat space.}}
		}
	
		\caption{The native and the tourist see the same cue stick
    	in the same hyperbolic space, yet they interpret what they see
    	very differently.}

	\label{Fig:MentalModels}
	\end{minipage}
\end{figure}

\noindent If, while you're playing billiards, you shut one eye, the effect disappears immediately: the cue stick instantly looks straight!  If you spent enough hours playing hyperbolic billiards (with both eyes open!), your brain might start to see the cue stick as straight, but I think that would be an unhealthy experiment. Please don't try it.

Exercise for the reader:  If a person born and raised in hyperbolic space were to come visit us here in Euclidean space, how would she perceive one of our cue sticks?

\end{quotation}

\section{Conclusions and future work}
\label{Sec:ConclusionsAndFutureWork}

We have noted that in any VR simulation of curved space, the geometrical effects of holonomy would cause the player's head and hands to gradually lose coherence, in the sense that what the player sees with her eyes would become inconsistent with what she feels with her hands.  To maintain body coherence, the simulation must compensate for those holonomy effects.  In the case of the curved-space billiards game, there's the additional constraint of maintaining coherence with a physical table.  Fortunately the table facilitates an effective solution:  we compute the player's head and hand poses relative to a nearby point on the table edge, which ensures that the head and hands stay coherent with the table as well as with each other.  In the case of VR simulations with no tangible components, we could either compute the hand poses relative to the head, or compute both relative to some virtual object that we expect to remain close to the center of the user's play area.  Future work here may depend heavily on the details of the individual VR applications.

We have found that the same computer code often works for all three isotropic spaces (spherical, Euclidean and hyperbolic), with almost no need to split into separate cases, if certain principles are followed.  The simplest---but most surprising---such principle is the need to assign a ``radius'' to a simulated Euclidean space (just as we assign a radius to a simulated 3\nobreakdash-sphere and to a simulated hyperbolic 3\nobreakdash-space), and to measure Euclidean translation distances in ``Euclidean radians'' (just as we measure rotations of a sphere in radians, and measure translations of hyperbolic 3\nobreakdash-space in hyperbolic radians).  A second principle is that a slight modification to how we visualize the projection transformation lets us easily see the spherical and hyperbolic analogs of the traditional Euclidean view frustum.  Those principles, together with a careful distinction between distances in meters and distances in radians, and some special attention to the back hemisphere of the 3\nobreakdash-sphere, allow for accurate and efficient rendering in the three isotropic spaces.

We have found that a Euclidean-born tourist visiting a curved space grossly misjudges distances, because when her eyes focus on an object, her brain misinterprets the vergence angle based on her prior Euclidean experience, rather than on the required curved-space relationship between vergence angle and distance.  To correct for this, we modified the graphics pipeline so that a Euclidean-born tourist will perceive each object at its correct curved-space distance.

I was surprised to discover that, even with the vergence-angle adjustment just described, a Euclidean-born tourist nevertheless perceives the space differently than a native-born inhabitant would perceive it.  The reason is that after one part of the player's brain interprets vergence angles as perceived distances, some other part of the player's brain integrates those perceived distances to form a mental model of the 3\nobreakdash-dimensional space.  Much work remains to be done to understand how people's mental model responds to being in a curved space and, for example, playing a game like billiards there.  Because such an investigation would extend far beyond my own skill set, I have made plans with a psychology professor (a specialist in spatial perception) for us to test this in her lab.  At the most basic level, it will be interesting to check whether all people perceive curved spaces the same way.  For example, for a given space (spherical, Euclidean or hyperbolic) and a given viewing mode (native-inhabitant or tourist), do all people agree about whether the billiards table and the cue stick appear convex, straight, or concave?  When they shut one eye, do they all agree that the cue stick instantly looks straight?  At a deeper level, I look forward to learning more about people's mental model of space, how it responds to curved-space experiences, and how it might vary from person to person.  Unfortunately the planned laboratory experiments cannot begin until after the COVID\nobreakdash-19 pandemic gets brought sufficiently under control.


\bibliography{BodyCoherenceRefs}

\end{document}